\documentclass[twocolumn,iop,numberedappendix,twocolappendix]{emulateapj}
\usepackage{graphicx}
\usepackage{xparse}
\usepackage{xcolor}
\usepackage{bm,upgreek}
\usepackage{amsmath}

\newcommand{\hi}{H\textsc{i}}
\newcommand{\kpar}{k_{||}}
\newcommand{\kperp}{k_\perp}

\newif\ifanalytic

\shorttitle{Baseline layouts for EoR wedge}
\shortauthors{Murray \& Trott}

\ExplSyntaxOn
\NewDocumentCommand\vect{m}
{
	\commexo_vector:n { #1 }
}

\cs_new_protected:Npn \commexo_vector:n #1
{
	\tl_map_inline:nn { #1 }
	{
		\commexo_vector_inner:n { ##1 }
	}
}

\cs_new_protected:Npn \commexo_vector_inner:n #1
{
	\tl_if_in:VnTF \g_commexo_latin_tl { #1 }
	{
		\mathbf { #1 } 
	}
	{
		\tl_if_in:VnTF \g_commexo_ucgreek_tl { #1 }
		{
			\bm { #1 } 
		}
		{
			\tl_if_in:VnTF \g_commexo_lcgreek_tl { #1 }
			{
				\commexo_makeboldupright:n { #1 }
			}
			{
				#1 
			}
		}
	}
}

\cs_new_protected:Npn \commexo_makeboldupright:n #1
{
	\bm { \use:c { up \cs_to_str:N #1 } }
}

\tl_new:N \g_commexo_latin_tl
\tl_new:N \g_commexo_ucgreek_tl
\tl_new:N \g_commexo_lcgreek_tl
\tl_gset:Nn \g_commexo_latin_tl
{
	ABCDEFGHIJKLMNOPQRSTUVWXYZ
	abcdefghijklmnopqrstuvwxyz
}
\tl_gset:Nn \g_commexo_ucgreek_tl
{
	\Gamma\Delta\Theta\Lambda\Pi\Sigma\Upsilon\Phi\Chi\Psi\Omega
}
\tl_gset:Nn \g_commexo_lcgreek_tl
{
	\alpha\beta\gamma\delta\epsilon\zeta\eta\theta\iota\kappa
	\lambda\mu\nu\xi\pi\rho\sigma\tau\upsilon\phi\chi\psi\omega
	\varepsilon\vartheta\varpi\varphi\varsigma\varrho
}

\ExplSyntaxOff

\begin{document}
	
	
	\title{The effect of baseline layouts on the EoR foreground wedge: a semi-analytical approach}
	
	\author{Steven G. Murray\altaffilmark{1,2,3,4} \and C.~M.~Trott\altaffilmark{2,3}}%
    
%
%
	
	
	\altaffiltext{1}{ARC Centre of Excellence for All-sky Astrophysics (CAASTRO) }%
	\altaffiltext{2}{International Centre for Radio Astronomy Research (ICRAR), Curtin University,  Bentley, WA 6102, Australia}
	\altaffiltext{3}{ARC Centre of Excellence for All-Sky Astrophysics in 3 Dimensions (ASTRO 3D)}
	\altaffiltext{4}{\email{steven.murray@curtin.edu.au}}

		
	
%






\begin{abstract}
The 2D power spectrum is a cornerstone of the modern toolkit for analysis of the low-frequency radio interferometric observations of the 21 cm signal arising from the early Universe. 
Its familiar form disentangles a great deal of systematic information concerning both the sky and telescope, and displays as a foreground-dominated ``brick'' and ``wedge'' on large line-of-sight scales, and a complementary ``window'' on smaller scales.
This paper builds on many previous works in the literature which seek to elucidate the varied instrumental and foreground factors which contribute to these familiar structures in the 2D power spectrum.
In particular, we consider the effects of $uv$-sampling on the emergence of the wedge.
Our results verify the expectation that arbitrarily dense instrument layouts in principal restore the missing information that leads to mode-mixing, and can therefore mitigate the wedge feature.
We derive rule-of-thumb estimates for the required baseline density for complete wedge mitigation, showing that these will be unachievable in practice.
We also discuss the optimal shape of the layout, showing that logarithmic regularity in the radial separation of baselines is favourable.
While complete suppression of foreground leakage into the wedge is practically unachievable, we find that designing layouts which promote radial density and regularity is able to reduce the amplitude of foreground power by 1-3 orders of magnitude.

\end{abstract}

\section{INTRODUCTION }
\label{sec:intro}

The 2D power spectrum (PS) has proven to be a valuable diagnostic quantity in the hunt for the signature of the Epoch of Reionisation (EoR) in the early \hi\ density field \citep{Liu2014,Trott2014,Barry2016}.
While the 21cm EoR signal is expected to be well-described by the 1D PS, whose interpretation assumes both homogeneity and isotropy in 3 dimensions,
its processing through low-frequency radio telescopes, as well as the various foreground contaminants, do not preserve 3D isotropy.
Indeed, radio interferometers are remarkably complex and exhibit different systematic effects in frequency and spatial modes, corresponding to line-of-sight ($\kpar$) and perpendicular ($k_\perp$) modes respectively.
The 2D PS maintains the separation of these modes, which is highly useful in identifying and removing/mitigating systematics.

It is well known that the dominant component of the raw data arises not from the desired 21 cm signal, but the extremely bright foregrounds \citep[eg.][]{Shaver1999}. 
Mitigation of these foregrounds has become one of the key areas of research in the field \citep[eg.][]{Bernardi2009,Liu2011,Chapman2012,Chapman2013,DiMatteo2004,Dillon2015,Chapman2015,Jensen2016,Sims2016,Mertens2017,Chapman2016,Ghosh2017,Murray2017}, and a standard picture has emerged which centres on the expected spectral smoothness of the foregrounds in contrast to the EoR signal. 
In this picture, the foregrounds are constrained to low-$\kpar$ modes in the 2D PS, leaving high-$\kpar$ modes relatively free of their contamination (for a clear schematic, see Fig. 1 of \citet{Liu2014a}).
Various methods have been developed to exploit this behaviour, either by entirely \textit{avoiding} low-$\kpar$ modes, \textit{subtracting} spectrally-smooth foreground models (either parametrically or non-parametrically), or a combination of the two \citep{Chapman2016}.

These methods have a well-known complication, however: any radio interferometer is by nature \textit{chromatic}, i.e. its view of the sky differs at different frequencies (via a number of factors including the frequency-dependent antenna response and the geometry of the baselines). 
This results in different \textit{perpendicular} scales having different dependency on frequency, i.e. \textit{parallel} scales -- a ``mode-mixing'' that can result in low-$\kpar$ power being ``leaked'' into higher $\kpar$ modes for higher $k_\perp$. 
This gives rise to the infamous foreground ``wedge'', and its complementary EoR ``window''.

The wedge causes issues for foreground mitigation techniques. 
For example, consider the ``avoidance'' technique: firstly, the existence of the wedge increases the number of modes over which the foregrounds dominate, and therefore further obscures the faint signal. 
Secondly, if one incorrectly specifies the boundary of the wedge, the extra leaked power inside the supposed ``window'' will at best bias results, if not overwhelm them.

The foreground wedge has been the subject of a great deal of research over the past decade.
It was originally reported by \citet{Datta2010}, though it was alluded to in different form in previous work \citep[eg.][]{Bowman2009}.
Following this initial report, a handful of papers appeared that included intuitive and pedagogical explanations for why the wedge arises, along with some discussion as to the primary physical factors which determine its shape and amplitude \citep{Morales2012,Parsons2012,Trott2012,Vedantham2012}.
These various explanations are complementary in the sense that they appraise the problem from different perspectives, arriving at similar conclusions.
For instance, \citet[hereafter M12]{Morales2012} explain the foreground wedge in the context of \textit{residual foreground emission} from point-sources (eg. arising from imperfect source peeling), considering as an example a single point-source at some location on the sky. 
They argue that the wedge arises due to the combination of the migration of baselines in the $uv$-plane with frequency, and the missing information inherent to a discrete sampling of the $uv$-plane. 
Figure 2 in M12 presents their intuitive picture: the inherent voltage correlations for a given baseline length are sampled by a single baseline which moves across the corrugated correlations at an angle in frequency. 
After accounting for smearing from the instrumental beam, the gridded baseline, when Fourier-Transformed over frequency, contains oscillations that introduce power at mode $\kpar > 0$, dependent on how quickly the baseline moves through the corrugations in frequency.
The value of $\kpar$ at which this power is introduced is simply proportional to the ``slope'' of the baseline in the $|u|\nu$-plane, which is itself proportional to the baseline length or $\kperp$. 
Furthermore, as sources closer to the horizon have higher oscillations in their voltage correlations as a function of baseline length, this mode-coupled $\kpar$ value is also proportional to the distance of the source from phase-centre: $\kpar \propto l \kperp$. 
Since sources are constrained to be within the horizon, we have $l_{\rm max} = \sin \theta_{k, {\rm max}} = 1$, and we are able to define the ``horizon limit'' beyond which we do not expect flat-spectrum foreground sources to contribute.
 
M12 also argues that the wedge is \textit{fundamental} in low-frequency interferometric observations, and cannot be avoided (at the relevant modes) merely by clever analysis, such as visibility gridding and weighting schemes.

\citet{Trott2012} (T12) reformulates the description of M12 in the context of a uniform distribution of faint undetected sources, deriving an exact analytical form  for the expected foreground power under a set of simplifications.

Alternatively, \citet[hereafter P12]{Parsons2012} describes the emergence of the wedge in terms of the so-called ``delay transform''. 
The delay transform considers a single baseline at a time, and associates `delay'-modes -- the fourier-dual of the frequency-track of the baseline -- with the line-of-sight modes $\kpar$. 
The delays themselves are simply time-delays between the reception of plane-wave emission at the two antennas composing the baseline (see Fig. 1 of P12 for a clear diagram).
In this scheme, the arguments are entirely geometric;
for a given source, not at phase centre, a longer baseline will correspond to a higher delay. 
Similarly, for a given baseline, a source closer to the horizon will correspond to a higher delay -- with a physical maximum at the horizon. 
This leads to the now familiar equation relating the line-of-sight mode at which power from a single source manifests: $\tau \sim \kpar \propto l \kperp$.
P12 explains that the delay transform ideally maps a flat-spectrum point source to a delta-function in delay space, but that in practice the non-flat spectral properties of both source and instrument add a (hopefully narrow) kernel which can throw power into modes beyond the ``horizon limit'' (cf. Fig. 1 of P12).
They thus suggest that designing instruments with maximal spectral and spatial smoothness, as well as reduced field-of-view, and then ignoring modes below the horizon limit, is a useful way to avoid the foreground problem. 
This has motivated the design of the PAPER \citep{,Ali2015} and HERA \citep{DeBoer2016} experiments. 

Despite the simplicity of these intuitive descriptions of the emergence of the wedge, its precise amplitude and shape are dependent on a combination of various complex effects. 
Amongst these are unavoidable sky-based effects such as the (spatially varying) spectrum of sources and diffuse emission, angular distribution of sources \citep{Bowman2009,Trott2016a,Murray2017}, effects of co-ordinate transformation from curved to flat sky \citep{Thyagarajan2013,Thyagarajan2015,Ghosh2017} and polarization leakage \citep{Gehlot2018}, as well as spectral characteristics of the instrument, such as the beam attenuation pattern, bandpass, chromatic baselines and chromatic calibration errors \citep{Bowman2009,Thyagarajan2015,Pober2015,Trott2016a}.
Due to the complexity of these effects, they are often investigated either by using simulations or via analytic simplifications which elucidate the effects of some subset of the components. 
Several works have been devoted to developing general frameworks to model the foreground wedge in order to mitigate it effectively \citep[eg.][]{Liu2014,Liu2014a,Pober2015,Ghosh2017}.

Despite the breadth of this research, one aspect which seems to have gained little attention is the layout of the antennas (or correspondingly the baselines) themselves,
and how they might be used to mitigate the wedge.
This is perhaps surprising as several of the seminal works on the topic suggest that one way to alleviate mode-mixing is to employ dense $uv$-sampling so that $uv$-samples overlap at various frequencies \citep[eg.][]{Bowman2009,Morales2012,Parsons2012}.
While a perfect $uv$-sampling is clearly unachievable, which has perhaps led to this avenue being largely ignored, it would seem advantageous to determine the extent of wedge-suppression possible under reasonable constraints. 

The purpose of this paper is to explore this question, which we attack in two parts. First we approach the question of how the wedge relates to the baseline layout, seeking intuitive semi-analytical understanding of the factors involved. Secondly, we ask the more pointed question of how far the wedge might be suppressed merely by choice of layout, limiting ourselves to layouts which might be realistically achieved. For this latter question, we necessarily turn to simple numerical simulations.
We approach the questions from a pedagogical view, making simplifications where necessary in order to elucidate conceptual understanding.




The layout of the paper is as follows. \S\ref{sec:framework} introduces the general equations (and assumptions) used throughout this paper to define the expected 2D PS, and the model simplifications we adopt. 
\S\ref{sec:classic} presents the classical form of the wedge by solving the general equation for a suitably sparse layout, which is shown to be equivalent to the delay spectrum.
This lays the groundwork for \S\ref{sec:weighted}, which considers the same family of $uv$-sampling functions, but with increased density, and thus must account for correlations between baselines.
It provides a semi-analytic framework to describe the density of baselines required to mitigate the wedge, and also the effects of deviations of the baseline layout from the assumed perfect regularity.
\S\ref{sec:mitigation:arrays} turns to discussion of the consequences of the preceding results for realistic arrays, and analyses explicit wedge reduction for a series of archetypal layouts.
Finally, \S\ref{sec:conclusions} wraps up with a summary of the key arguments and conclusions throughout the paper, and a prospectus for future work.

\section{Framework for Expected Foreground Power}
\label{sec:framework}
In this section we derive a (simplified) general equation that describes the \textit{expected} 2D PS for a given sky distribution and instrument model, following similar lines as \citet{Trott2012}, \citet{Liu2014} and \citet{Trott2016}\footnote{Readers familiar with these derivations should be able to skim this section lightly, referring to Eqs. \ref{eq:simple_vis}, \ref{eq:vis_gridded} and \ref{eq:power_general} and Tables \ref{tab:assumptions} and \ref{tab:models} thereafter.}.
We differ from \citet{Liu2014} in that we express the expected power (closely related to the covariance of visibilities) in the basis of the natural coordinates, ($\vect{u}$, $\eta$), rather than baseline vectors and delay (they express their covariance in the latter basis, and reconstitute in cosmologically aligned coordinates via another transformation).
This makes sense for our analysis, as we are interested in the conceptual understanding of where foreground power arises from, rather than a numerically efficient power spectrum estimator.
We also differ from \citet{Trott2016} in that we consider correlations between baselines in the expectation of the foreground power, which is necessary in order to properly evaluate the effects of $uv$-sampling.  

We \textit{a priori} remark that our framework is not fully general -- it does not include all possible factors.
This is in the hope of elucidating our primary goal -- the effect of the antenna layout.
One simplification we will enforce for this paper is that we only consider the effect of \textit{point sources}, not Galactic emission, or extended sources (or the negligible EoR signal for that matter). 
The extension to these other sources of foreground emission is neither conceptually important for this work nor conceptually difficult (though the details of the formulation can be rather involved, eg. \citet{Trott2017,Murray2017}).
A second simplification is that we consider the simple case in which the telescope is pointing instantaneously at zenith.
This alleviates complications arising from baseline foreshortening (not entirely, though enough for the conceptual understanding aimed for in this work, cf. \citet{Thyagarajan2015}).
Further, for simplicity, we will assume a perfectly co-planar array. 
The effect of relaxing these assumptions is expected to modulate power within the wedge, and potentially extend its reach to some degree.
Nevertheless, our focus is on examining the fundamental reason for the wedge, and whether it may be suppressed via appropriate $uv$-sampling -- thus focusing on a simple subset is appropriate. 
We attempt to exhaustively list the various global assumptions and simplifications we have made in Table \ref{tab:assumptions}.
We will outline further model simplifications and choices as we develop the equations within this section.

\begin{table*}[!ht]
	\begin{center}
		\begin{tabular}{ l }
			\hline
			\textbf{Assumptions used in framework} \\ 
			\hline
			Restriction to extra-galactic point sources \\
			Zenith-pointing only \\
			Co-planar antenna array \\
			Flat-sky approximation \\
			Naturally-weighted baselines \\
			Thermal noise values of all baselines drawn from i.i.d Normal distribution, centred on zero \\
			Flat-spectrum sources \\
			\hline
			\hline	
		\end{tabular}
	\end{center}
\caption{Summary of universal assumptions and simplifications used in this paper. \label{tab:assumptions}}
\end{table*}

\subsection{Single-baseline visibility}
We begin with the simple visibility equation for a co-planar array, which defines the signal received by any baseline, in the presence of point sources and thermal noise:
\begin{equation}
\label{eq:simple_vis}
V_{i}(\nu, \vect{u}_i) = \phi_\nu \left[\mathcal{N}_{i,\nu} + \int d\vect{l} dS  \   n(\vect{l}, S) SB(\nu, \vect{l}) e^{-2\pi i \vect{u}_i\cdot \vect{l}}\right],
\end{equation}
i.e. the Fourier-transform over the sky of the emission brightness attenuated by the beam $B$ and a frequency taper $\phi_\nu$.

\begin{figure}
	\centering
	\includegraphics[width=1.0\linewidth]{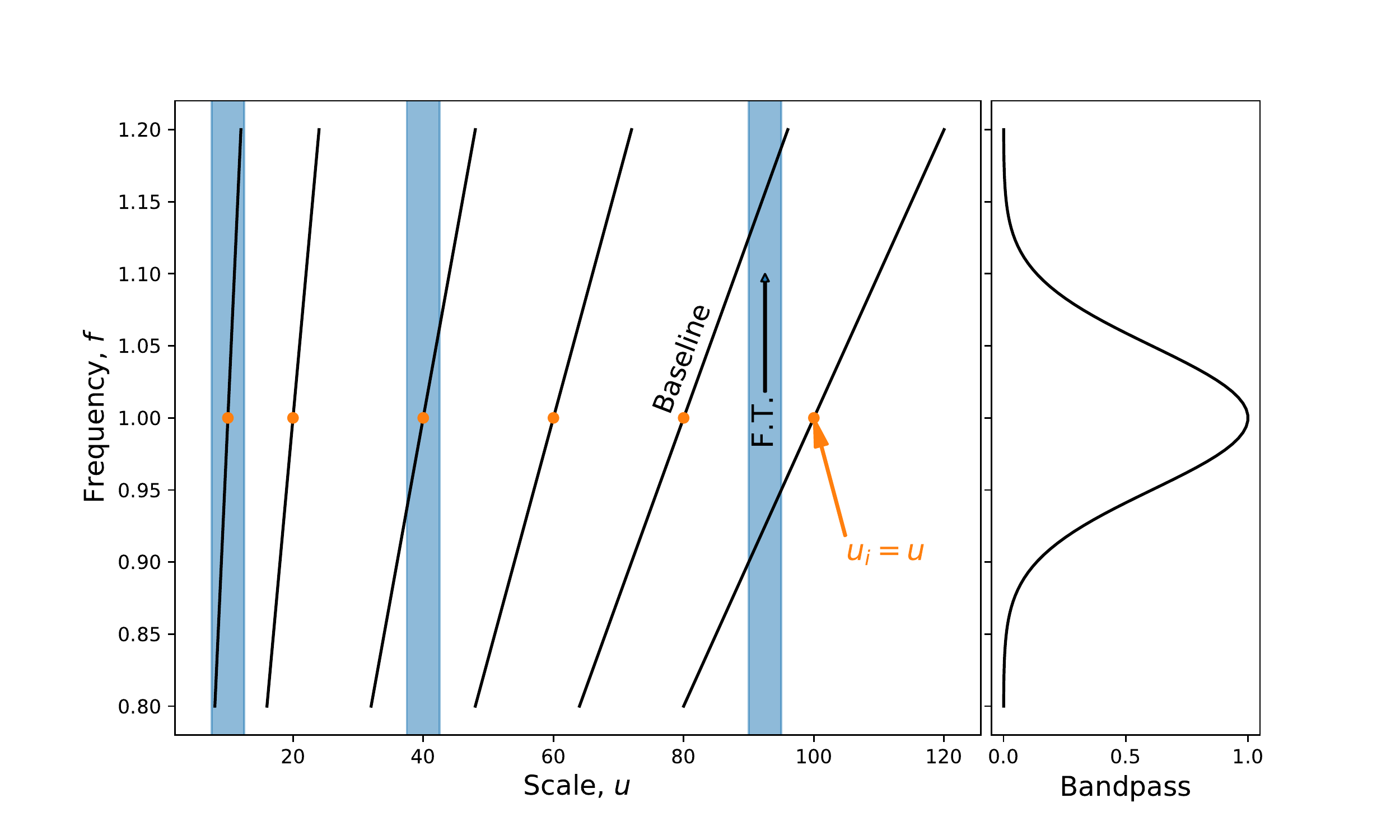}
	\caption{Schematic showing the migration of baseline length (in wavelength units) as a function of frequency. The vertical shaded regions indicate the axis of the frequency Fourier Transform, showing that multiple baselines contribute at different frequencies for high $u$. Orange dots show where each baseline is equivalent to the scale at which it would be equated in the delay approximation. At low $u$, a FT along a baseline is almost equivalent to the true FT, showing the delay approximation to be accurate.}
	\label{fig:baseline_schematic}
\end{figure}

The vector $\vect{u}_i$ is the baseline length in units of the observational wavelength:
\begin{equation}
\vect{u}_i = \vect{b}_i/\lambda,
\end{equation}
wherein lies the chromaticity of the $uv$-sampling.
Henceforth,
we let $\nu_0$ be a reference frequency (this will later be tied to the mid-point of the frequency band of observation without loss of generality), and define $f = \nu/\nu_0$.
We also explicitly let $\vect{u}_i$ denote the value of $\vect{u}_i$ \textit{at the reference frequency}.
This is illustrated in Fig.~\ref{fig:baseline_schematic}, which also shows why the delay approximation -- identifying the vertical shaded regions with the diagonal baseline tracks -- is reasonable for small $u$. 

In this paper, we will exclusively use a Gaussian-shaped beam, and predominantly it will be achromatic\footnote{For demonstration purposes, \S\ref{sec:classic} will also use a chromatic Gaussian beam, for which we have
	\begin{equation}
	\sigma_\nu = 0.42c/\nu D = \sigma_0/f. 
	\end{equation}
}. 
The Gaussian beam is thus
\begin{equation}
	B_\nu(l) = e^{-l^2/2\sigma^2},
\end{equation}
where $\sigma = 0.42c/\nu_0 D$ is the beam width \citep{Trott2016}.
The motivation for using a Gaussian beam is that it is the most realistic analytically-tractable form possible, and its use for conceptual studies has precedent \citep{Liu2014}.
We do note that the choice of a smooth Gaussian, which does not have sidelobes, has desirable effects on the form of the wedge, in that it perfectly suppresses horizon sources. 
This effectively combats the complexities of baseline foreshortening at the horizon due to our appropriated flat-sky approximation.

The emission brightness is written in Eq.~\ref{eq:simple_vis} as the sum of the flux density of all point-sources in the sky, where the number counts of these sources are given by $n(\vect{l},S)$. These differential number counts are in general a statistical quantity, as we will typically consider sources below the confusion limit of an instrument. 
To simplify the calculations to follow, we have followed common pedagogical practice and assumed that the the spectral shape of each source is entirely flat. 
This simplification is rather heavy, but it should not affect the \textit{conceptual} understanding of the following calculations.

Furthermore, the observed frequency window is both physically attenuated by the instrument, and will also be tapered within the analysis in order to suppress frequency side-lobes\footnote{Note that each sub-band also has its own structure. This may be assumed to be a part of $\phi$, or may be introduced as a secondary convolution \citep[cf. $\gamma$ in][]{Liu2014}. We shall ignore it in this work so as not to complicate the key ideas.}. 
In this paper, we assume that the bandpass is relatively broad compared to the taper, and that its effect can be safely ignored\footnote{We note that this is a particularly strong simplification. 
	The bandpass will in general introduce smaller-scale oscillations into $\phi$, which tend to broaden the footprint of the foreground power in $\omega$. This is in some way countered by our use of a broad $\omega$-space Gaussian taper.}.
We normalize the frequency-taper to $\phi(0) = 1$ 
\footnote{This assumption, which we employ for simplicity throughout this paper, ties the central frequency of the bandpass to the reference frequency, $\nu_0$. 
This is without loss of generality, as we may always shift all frequency-dependent parameters to a new ``reference'' before any analysis.}, and exclusively use a Gaussian taper here for tractibility:
\begin{equation}
	\phi(f-1) = e^{-\tau^2 (f-1)^2},
\end{equation}
with $\tau$ an inverse-width (or precision).

More common choices for the taper are the Blackman-Harris \citep{Trott2016a} or its self-convolution \citep{Thyagarajan2016}.
These serve to reduce leakage of power into higher modes, and are better choices than a Gaussian in practice. 
We utilise the Gaussian for analytic simplicity and note that most qualitative conclusions of the paper are insensitive to this choice (note also that there is precedent for choosing such a taper for theoretical studies, in \citet{Liu2014}).
The primary point of difference is in the definition of the ``brick'' (cf. Table \ref{tab:simple_summary}), which extends to a higher value of $\omega$ when using a Gaussian.
This also affects the position of the emergence of the wedge.

Taking the Fourier transform (over $\nu$), we arrive at
\begin{align}
	\label{eq:delay_vis}
	\tilde{V}_i(\eta, \vect{u}_i) = \nu_0 & \int df\ \ e^{-2\pi i \nu_0f \eta} V_i(\nu, \vect{u}_i).
\end{align}
Note that we will make the change of variables $\omega = \nu_0 \eta$ for the remainder of this work, where $\omega$ is dimensionless and makes for simpler theoretical equations.
The square of this particular quantity is called the \textit{delay spectrum}, and we explore it briefly in \S\ref{sec:classic}.

\subsection{Multi-baseline visibility}
To increase signal-to-noise, we typically \textit{grid} the discrete baselines in some fashion.
The delay spectrum approach only grids measurements after squaring the visibilities, whereas image-based approaches grid the complex visibilities before squaring to form the power.
We follow the latter approach in this work, as it is required in order to utilise the benefits of the $uv$-sampling.
However, the approaches can be forced to align with each other under special array layout conditions, and we explore this briefly in \S\ref{sec:classic}.

In essence, the gridding assigns a weight to each baseline, specifying its contribution to a given UV point (i.e. closer points receive higher weights) 
\footnote{We are also free to choose a baseline weighting which is a function only of the magnitude of $u_i$. Nevertheless, previous work has shown that an optimal choice for the baseline weighting is to have each baseline weighted equally \citep{Bowman2009,Parsons2012}, and we follow suit here.}. 
The total estimated visibility at point $\vect{u}$ is thus the sum of all weighted visibilities, normalised by the total weight.
Letting $w_\nu$ denote the weighting function, and defining the total weight as
\begin{equation}
\label{eq:wnuu}
W_\nu(\vect{u}) = \sum_{i=1}^{N_{\rm bl}}w_\nu(\vect{u} - f \vect{u}_i),
\end{equation}
we have
\begin{equation}
\label{eq:vis_gridded}
V(\nu, \vect{u}) =  \frac{1}{W_\nu(\vect{u})}\sum_{i=1}^{N_{\rm bl}} w_\nu(\vect{u} - f\vect{u}_i) V_i(\nu,\vect{u}_i).
\end{equation}
The Fourier-space visibility is then
\begin{align}
	\label{eq:vis_full}
	\tilde{V}(\omega, \vect{u}) = \nu_0 & \int df\ \ e^{-2\pi i f\omega} V(\nu, \vect{u}),
\end{align}
which may be squared to form the grid-based PS.

The combination of Eqs. \ref{eq:simple_vis}, \ref{eq:vis_gridded} and \ref{eq:vis_full}  provide the basis for all following work.

It has been shown in T16 that an unbiased gridding of visibilities is determined by inverting a matrix involving the fourier-transformed primary beam\footnote{In their work (cf. their Eq. 17) it also involves a matrix $\vect{G}$ which accounts for sky-curvature and other effects which we ignore here.}.
Indeed, they find that a good approximation to this matrix inversion, which enhances computability considerably, is to use the diagonalized inversion, which corresponds precisely to a weighted average of baselines, with a weighting function given by the fourier-transform of the beam, $\tilde{B}(\vect{u})$. 

Explicitly, for the Gaussian beam employed in this work, we have
\begin{equation}
w_\nu(u) =  B(u) = e^{-2\pi^2 \sigma^2 u^2}.
\end{equation}


\subsection{Statistical properties of the visibility}
The visibility is in general a statistical variable, both because of the random thermal noise and the (typically) statistical nature of $n(\vect{l},S)$.
$\tilde{V}$ is in general not Gaussian, nevertheless as we will be dealing with the power spectrum -- a quadratic quantity -- we shall only be required to know up to second-order properties of $\tilde{V}$ for the purposes of this paper.
As long as the thermal noise is independent of the foreground signal, these are simply derived.

Our primary sky model consists of a uniform Poisson process of point sources (uniform in $\vect{l}$, cf. T16)\footnote{This is a reasonable approximation for a relatively narrow beam (with no side-lobes) at zenith, where $\theta \sim \vect{l}$. Our adoption of the Gaussian beam means this assumption will have little consequence. In reality, the curved nature of the sky introduces excess brightness towards the horizon, which can result in the ``pitchfork'' structure reported in \citet{Presley2015,Thyagarajan2015,Thyagarajan2016} (see also \citet{Kohn2016,Kohn2018})}.
In this model, expectation of the frequency-space visibility is obtained simply by replacing the sky emission with the mean flux density,
\begin{align}
	\label{eq:su_meanvis}
	\langle V_{i}\rangle(\nu, \vect{u}_i) &= \phi_\nu \int d\vect{l}\   \bar{S} B_\nu(\vect{l}) e^{-2\pi i f \vect{u}_i \cdot \vect{l}}. \nonumber \\
	&= \phi_\nu \bar{S} B(f\vect{u}_i),
\end{align}
and the expected Fourier-space visibility is
\begin{align}
	\langle V (\omega, \vect{u})\rangle = \bar{S} \nu_0 \int df \frac{e^{-2\pi i f \omega}\phi_{\nu}}{W_\nu(\vect{u})} \sum_{i=1}^{N_{\rm bl}} B(\vect{u} - f\vect{u}_i)  B(fu_i). 
\end{align}

The variance of the visibility is composed of two parts -- a thermal variance and sky variance -- which are assumed to be independent. 
The sky variance may be worked out in similar fashion to the procedure outlined in \cite{Murray2017}.
The total covariance (i.e. with thermal noise term) is then
\begin{align}
\label{eq:su_covvis}
\vect{C}_{\rm bl} &=  \phi_\nu^2 \sigma_N^2 \delta_{ij} \delta(\nu-\nu')  \nonumber \\
& + \phi_\nu^2 \mu_2  \int d\vect{l}_1 B_{\nu}B_{\nu'} e^{2\pi i \vect{l} (f' \vect{u}_j  - f \vect{u}_i)},
\end{align}
between baselines $i$ and $j$, and frequencies $\nu, \nu'$, where we have used the assumed flat-spectrum of all sources, and $\mu_2$ is the second moment of the source count distribution:
\begin{equation}
\mu_n = \int dS\ S^n \frac{dN}{dS}.
\end{equation}
Thus the variance of the Fourier-space gridded visibility is
\begin{align}
	\label{eq:general_var}
	{\rm Var}(\tilde{V}) &=& \nu_0^2 &\int df df' \frac{e^{-2\pi i \omega (f-f')}}{W(\vect{u}) W'(\vect{u})} \nonumber \\
	&& & \times \sum_{ij} ^{N_{\rm bl}} B(\vect{u} - f\vect{u}_i)B(\vect{u} - f'\vect{u}_j) \vect{C}_{\rm bl} \nonumber \\
	&=& \nu_0^2 \sigma_N^2 &\int df\frac{ \phi_\nu^2 }{W^2_\nu(\vect{u})}  \sum_i^{N_{\rm bl}} B^2(\vect{u}-f\vect{u}_i) \nonumber \\
	& &\ +  \nu_0^2 &\int d\vect{l} df df' \frac{e^{-2\pi i \omega (f-f')}}{W_\nu(\vect{u}) W'_\nu(\vect{u})} \nonumber \\
	& & & \times \sum_{ij} ^{N_{\rm bl}} B(\vect{u} - f\vect{u}_i)B(\vect{u} - f'\vect{u}_j) \vect{C}_{\rm sky}.
\end{align}

The first term of this equation defines the thermal noise variance of a grid-point $(\omega, \vect{u})$. 
We omit the term for all following calculations\footnote{This can be thought of as going to the limit of a perfectly calibrated instrument, or infinite integration time, and thus setting $\sigma_N \rightarrow 0$.}, but we define
\begin{equation}
	\label{eq:w_theta}
	W^2(\vect{u}) = \int df \frac{ \phi_\nu^2 }{W^2_\nu(\vect{u})}  \sum_i^{N_{\rm bl}} B^2(\vect{u}-f\vect{u}_i)
\end{equation}
which will be used to determine a grid-point's relative weight when averaged to form the 2D PS.

\subsection{Expected power spectrum}
Armed with the statistical descriptors of the visibility, we can determine the expected 3D PS:
\begin{equation}
	\label{eq:expectation_equation}
	\langle P(\omega, \vect{u})\rangle  \equiv \langle \tilde{V}\tilde{V}^*\rangle = {\rm Var}(\tilde{V}) + |\langle \tilde{V} \rangle|^2.
\end{equation}
Note that for the statistically uniform sky that we primarily adopt (where uniformity is in $\vect{l}$), the second term is effectively the transfer function of the instrument (i.e. the beam and bandpass), and can typically be neglected on small perpendicular scales when these characteristics are angularly smooth.

The 2D PS is given by
\begin{equation}
	\label{eq:power_general}
	\langle P(\omega, u) \rangle = \frac{\int_0^{2\pi} d\theta \ \langle P(\omega, \vect{u})\rangle W^2(\vect{u})}{\int_0^{2\pi} d\theta \ W^2(\vect{u})} , 
\end{equation}
where $\theta$ is the polar angle of $\vect{u}$.

We provide a synopsis of the meaning of symbols used in this paper in Table \ref{tab:models}.

\begin{table*}[!ht]
	
	\begin{center}
		\begin{tabular}{ l l l }
			\hline
			\textbf{Symbol} & \textbf{Description} & \textbf{Models/Values}  \\ 
			\hline
			$\nu$ & Frequency & \\
			$f$ & Normalised Frequency, $\nu/\nu_0$ & \\
			$\vect{l}$ & Cosine-angle of sky co-ordinate, $\cos\theta$ & \\
			$\vect{b}$ & Baseline length & \\
			$\vect{u}$ & Fourier-dual of $\vect{l}$, equivalent to $\vect{b}/\lambda$ & \\ 
			$\eta$, $\omega$ & Fourier-dual (and scaled by $\nu_0$) of $\nu$ & \\
			$k_\perp$, $\kpar$ & Cosmologically-scaled $u$ and $\eta$ respectively & \\
			$V$ & Interferometric Visibility as function of frequency& \\
			$\tilde{V}$ & Frequency FT of $V$ & \\
			$S$ & Flux density (subscripted for a particular source) & \\
			$I(\nu,\vect{l})$ & Sky Intensity &  \\
			
			\hline
			$\mu_1 \equiv \bar{S}$ & Mean brightness of sky  & 1 Jy/sr \\
			$\mu_2$ & Second moment of source-count distribution & 1 Jy$^2$ /sr \\
			$S_0$ & Flux density of single source in sky & 1 Jy \\
			$\vect{l}_0$ & Position of single source in sky & (1,0) \\
			$\nu_0$ & Reference frequency & 150 MHz \\
			$\sigma$ & Beam-width at $\nu_0$ & 0.2 rad \\
			$\tau$ & Unitless band-pass precision, $1/2\sigma_f^2$ & 100 \\
			$D$ & Tile Diameter & 4m \\
			$\phi$ & Frequency Taper & Gaussian \\
			$\phi_B$ & Bandpass & Uniform \\
			$\psi$ & Source spectral shape & Flat \\
			\hline
			$B_\nu$ & Beam Attenuation & Gaussian (Static; Chromatic)  \\  
			$w$ & Visibility-gridding weights & Fourier-beam kernel \\
			$n(\vect{l},S)$ & Source count distribution & Single Source; Stochastic Uniform \\
			\hline
			\hline
			
		\end{tabular}
			\caption{Summary of symbols and models used throughout this paper. Where possible, parameters list their default value used in plots throughout this paper. Any models list all models explored throughout this paper.  Any Latin-subscripted perpendicular scale (eg. $\vect{u}_i$) refers to a particular physical baseline at reference frequency. \label{tab:models}}
	\end{center}
\end{table*}

\section{Uncorrelated Visibilities}
\label{sec:classic}
The wedge has been shown to naturally arise in the expected PS when correlations between baselines are either ignored or absent \citep[eg.][]{Parsons2012,Trott2016}.
For example the delay spectrum, which by definition cannot correlate visibility pairs, can be used to provide a simple illustration of the emergence of the wedge. 
To provide a backdrop for discussion of the effects correlating close baselines, we first turn to this uncorrelated case (within the context of our framework) to illustrate the emergence of the wedge.



In order to obtain simple single-baseline measurements of the PS within our framework, we use three conditions: i) baselines sparsely arranged on a set of logarithmic spokes, ii) an artificially narrow gridding kernel, and iii) for simplicity, we evaluate the PS only at reference baseline positions (i.e. $\vect{u} = \vect{u}_i$).

As we shall see, condition (iii) is not really required, but does simplify the procedure slightly.
Condition (ii) can be more precisely stated as setting the gridding kernel width to approach zero.
This is artificial, because we will not enforce the beam width to follow the same limit.
Alternatively, one may imagine condition (ii) as employing a nearest-baseline weighting method, such that the closest baseline to a point $\vect{u}$ at a given frequency will be the sole contributor.
We shall see that even for standard gridding kernels, this condition will be met for large $u$ if condition (i) is met.

\begin{figure}
	\centering
	\includegraphics[width=1.0\linewidth]{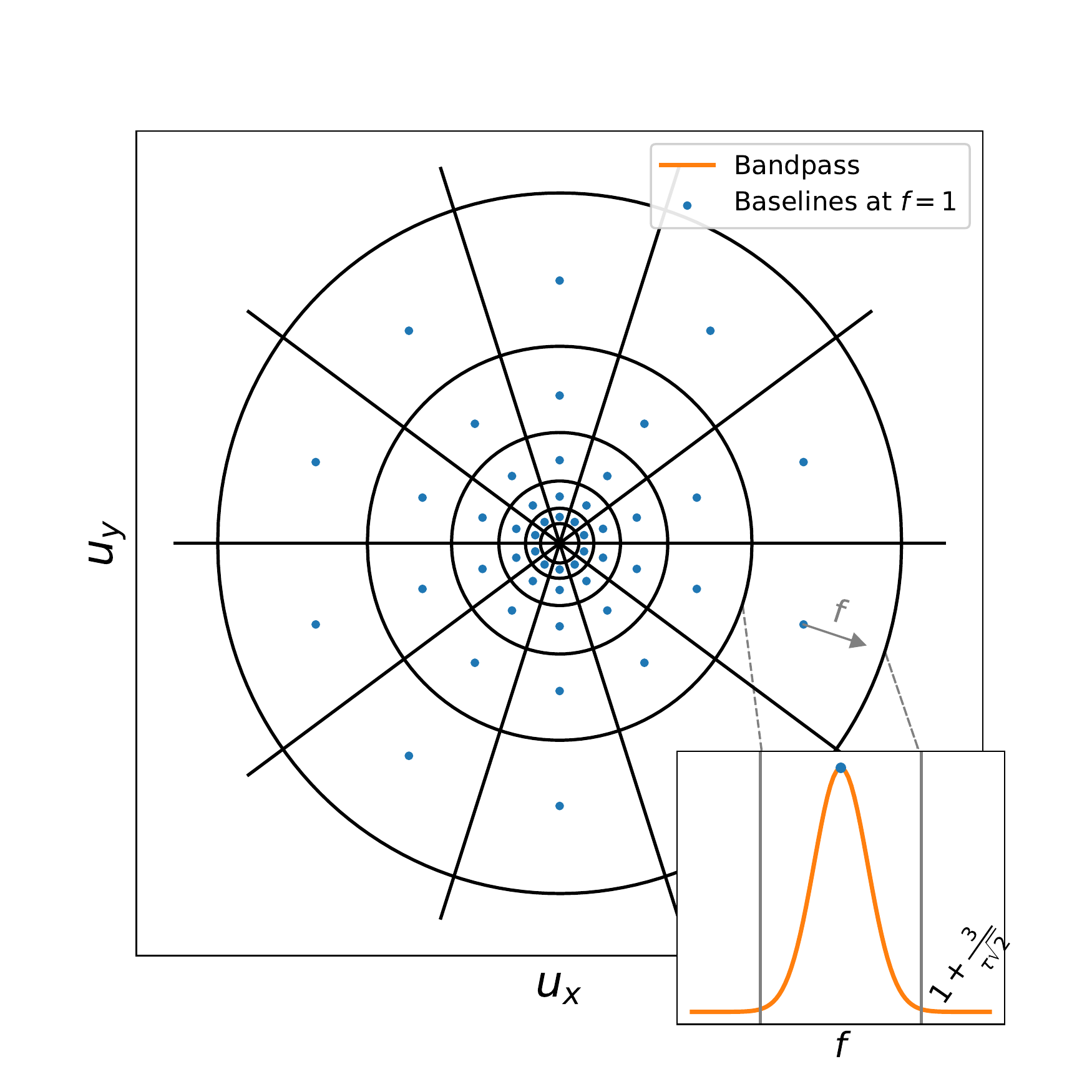}
	\caption{Schematic of baseline layout with overlaid averaging/histogram grid. Each cell of the averaging grid contains a single baseline which defines the "position", $\vect{u}_\mu$ of the cell at $f=1$. Inset is a figure transforming to frequency space, in which the bandpass/taper is shown, along with the frequencies at which a particular baseline is at the cell edges and centre. The edges correspond to 3 Gaussian widths of the bandpass.}
	\label{fig:delay_transform_schematic}
\end{figure}

Condition (i) is illustrated in Fig.~\ref{fig:delay_transform_schematic}.
The baselines are arranged in a regular (logarithmic) polar grid, or equivalently, a series of ``spokes" along which baselines are strung in a logarithmically regular fashion. 
Importantly for this section, the base of the logarithm is large enough such that a single baseline remains the sole contributor to a point co-located with its reference co-ordinate for the entire bandwidth (illustrated by the inset orange bell-curve in Fig.~\ref{fig:delay_transform_schematic}).
In addition, every spoke is equivalent, such that the baselines define a set of concentric rings. 
In summary then, the layout consists of $N_\theta \times N_r$ baselines, with regularly-spaced angular coordinates $\theta_k = 2\pi k/N_\theta$ and log-spaced radial co-ordinates $u_j = u_{j-1} + \Delta_u (u_{j-1})$, with $\Delta_u(u) = u\Delta$ (with constant $\Delta$) and arbitrary $u_0$.
We shall re-use variants of this simple layout throughout this paper.



These conditions allow for simple analytic solutions of the expected PS for a range of combinations of sky models and beam shapes.
While emergence of the wedge can in principal be illustrated without specialising to any particular model, we find it illustrative to do so.
Furthermore, to demonstrate that the wedge arises irrespective of sky model or beam shape, we use two models for each and evaluate the expected PS for three combinations: single-source with static beam, single-source with chromatic beam, and uniform sky with static beam.
Thus in addition to our fiducial choice of static Gaussian beam and uniform point-source sky, we also use a chromatic Gaussian beam (cf. \S\ref{sec:framework}) and a single-source sky model. 
The latter model places a single source of flux density $S_0$ at position $\vect{l}_0$.

The solutions are derived in App.~\ref{app:delay}, and shown in tabulated form in Table \ref{tab:simple_summary}. 
For completeness we show the assumed values of various parameters in Table \ref{tab:models}, and these correspond to the plots in Fig. \ref{fig:ss_ngp_static}.

\subsection{Discussion of Uncorrelated Examples}
The basic form of the three derived solutions is very similar, as can be most clearly seen in the final two columns of Table \ref{tab:simple_summary}, which provide a schematic view of where the power cuts off in $\omega$. 
In each case the standard form of the wedge, in which the power cut-off traces $\omega \propto u$, is recovered.
The constant of proportionality here (for this choice of coordinates) is given either by $l_0$ in the case of a single source (two leftmost plots, first with $l=1$ and second with $l=0.5$), or the beam width, $\sigma$ (rightmost plot), in the case of a stochastic uniform sky.
In either case, the maximum possible value for these is unity (in fact, for the beam width it must be less than this or else sky-curvature terms become important).
Thus we can draw the standard ``horizon line'' at $\omega = u$. 

We also find that the lower-left portion of the $\omega u$-plane forms a ``brick'' whose width is determined primarily --- in our case --- by the taper.
Clearly, the ``brick'' is in general determined by the overall frequency envelope of the instrument and analysis -- i.e. the combination of taper, bandpass, chromaticity of the beam, and spectral structure of the sources. 
Indeed, the effects of the chromaticity of the beam are apparent in the case of a single-source sky; the width of the brick is determined by $p^2 = \tau^2 + l_0^2/2\sigma^2$, which has a dependence on the beam-width. 
In practice, the taper/bandpass dominate the spectral response, and $p^2 \approx \tau^2$, nevertheless this illustrates that even with a theoretically infinite uniform bandpass, other spectral characteristics of the instrument will limit the ability to sequester power into the lowest $\omega$ modes. 
In general, the balancing of the various spectral terms provides the motivation for design criteria on the spectral smoothness of the instrument. 

As has been previously noted, the wedge occupies a greater portion of the $\omega u$-plane for sources close to the horizon (precisely because their delay transform for the same baseline is larger). 
Thus, a beam which is tighter (and doesn't have high-amplitude sidelobes) can effectively attenuate these sources and ameliorate the wedge \citep{Parsons2012}.
Of course, such a beam will also attenuate the 21 cm signal, and is therefore not an ideal solution for the problem. 

\begin{figure*}
	\centering
	\includegraphics[width=\linewidth,trim=2cm 0cm 2cm 0cm]{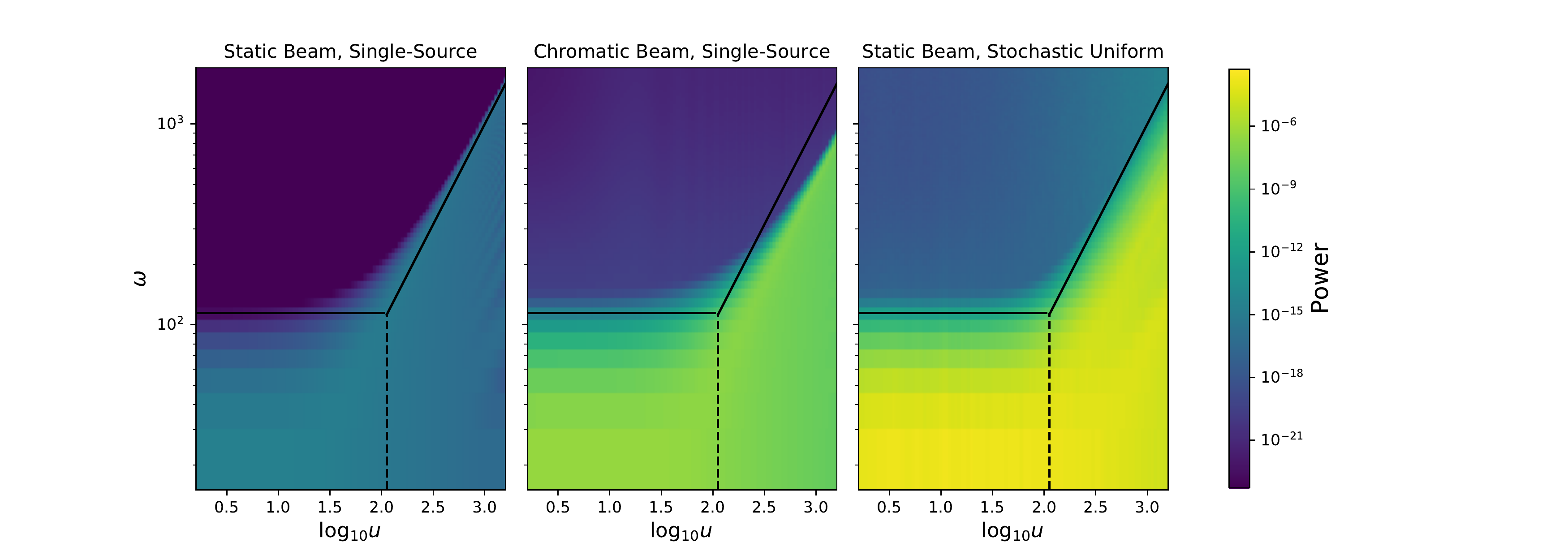}
	\caption{2D PS examples using averaged gridding, and different combinations of beam and sky models. Each displays very similar behaviour, however the wedge is
		sharper in the case of a single source. In each case, a horizontal ``brick'' line is drawn at the theoretical $10^{th}$ magnitude of suppression (cf. Table \ref{tab:simple_summary}), and the diagonal line is the ``horizon line'': $\omega = u$. For each, $\tau$ remains the same, while the two single-source models have sources at different zenith angles, the first at the horizon and the second at $l=0.5$. The clear difference in amplitude arises due to the difference in number of position of sources in the beam. Changing the position of a single source changes height and slope of the wedge line. A stochastic uniform sky has a softer edge for the wedge.}
	\label{fig:ss_ngp_static}
\end{figure*}

As has been extensively noted in the literature, many factors affect the precise form of the power in the wedge \citep[eg.][]{Thyagarajan2013,Thyagarajan2015,Thyagarajan2016,Gehlot2018} --- and most of these we have ignored in this analysis.
While the broad structure remains the same -- a brick with width given by the spectral envelope of the instrument, and linear wedge extending to $\omega \approx u$ -- the power within this region may be shifted around or amplified by various factors, and even leaked beyond the horizon line when small-scale spectral features are present in the analysis. 
An example of this changing of form can be witnessed in Fig.~\ref{fig:ss_ngp_static} between the single-source and stochastic skies. 
The smooth attenuation of sources at larger angles causes a smoother cut-off in the wedge.

Nevertheless, none of these features can lay claim to being the fundamental reason for the wedge. 
Altering them merely alters the shape or amplitude of the wedge, and not its basic form or existence.
The fundamental reason for the wedge is rather the combination of the migration of the baselines with frequency, and the sparsity of the $uv$-sampling. 
We turn to examining this latter condition for the remainder of this paper.


\begin{table*}
	\begin{center}
		\begin{tabular}{ l l l l l }
			\hline
			\textbf{Sky Dist.} & \textbf{Beam} & \textbf{Form of $P(\omega, u_\mu)$} & \textbf{Low $u$} &\textbf{High $u$}  \\ 
			\hline
			Single-Source & Static & \( \displaystyle \frac{1}{N_\theta}  \frac{S_0^2 \nu_0^2 \pi}{\tau^2} \exp\left(-\frac{l_0^2}{\sigma^2}\right) \sum_{k=1}^{N_\theta} \exp\left( -\frac{2\pi^2(\omega + u_\mu l_0 \cos (2\pi k/N_\theta))^2}{\tau^2} \right) \) &$\tau/\pi\sqrt{2}$ & $ul_0$ \\
			Single-Source & Chromatic & \(\displaystyle \frac{S_0^2 \nu_0^2 \pi}{N_\theta p^2} e^{-\tau^2l_0^2/2\sigma^2p^2} \sum_{k=1}^{N_\theta} \exp\left( -\frac{2\pi^2 (\omega + u l_0 \cos (2\pi k/N_\theta))^2}{p^2}\right) \) &$p/\pi\sqrt{2}$ & $ul_0$ \\
			Stochastic Uniform & Static & \(\displaystyle \frac{\nu_0^2 \pi}{p_u}\exp\left(-\frac{2\pi^2\omega^2}{p_u^2}\right) \left[ \frac{\bar{S}^2}{p_u} \exp\left(-\frac{2\tau^2\pi^2\sigma^2u^2}{p_u^2}\right) + \frac{\mu_2 \pi^2\sigma^2}{\tau}\right]\) & $\tau/\pi\sqrt{2}$ &  $u\sigma$ \\
			\hline
			\hline
		\end{tabular}
		\caption{\label{tab:simple_summary} Summary of analytic solutions for the simple discrete polar grid layout with histogram gridding of \S\ref{sec:classic}. Final two columns display a schematic representation of where the foreground power cuts off in $\omega$. For simplicity we list the cutoff such that the power is reduced by a factor of $e$ from the total. For a $\chi$-order of magnitude suppression, multiply the result by $\chi \ln 10$. In the table, $p_u^2 = \tau^2 + 2\pi^2\sigma^2u^2$ and $p^2 = \tau^2 + l_0^2/2\sigma^2$.}
	\end{center}
\end{table*}

\section{Dense Logarithmic Polar Grid}
\label{sec:weighted}

\begin{figure}
	\centering
	\includegraphics[width=1\linewidth]{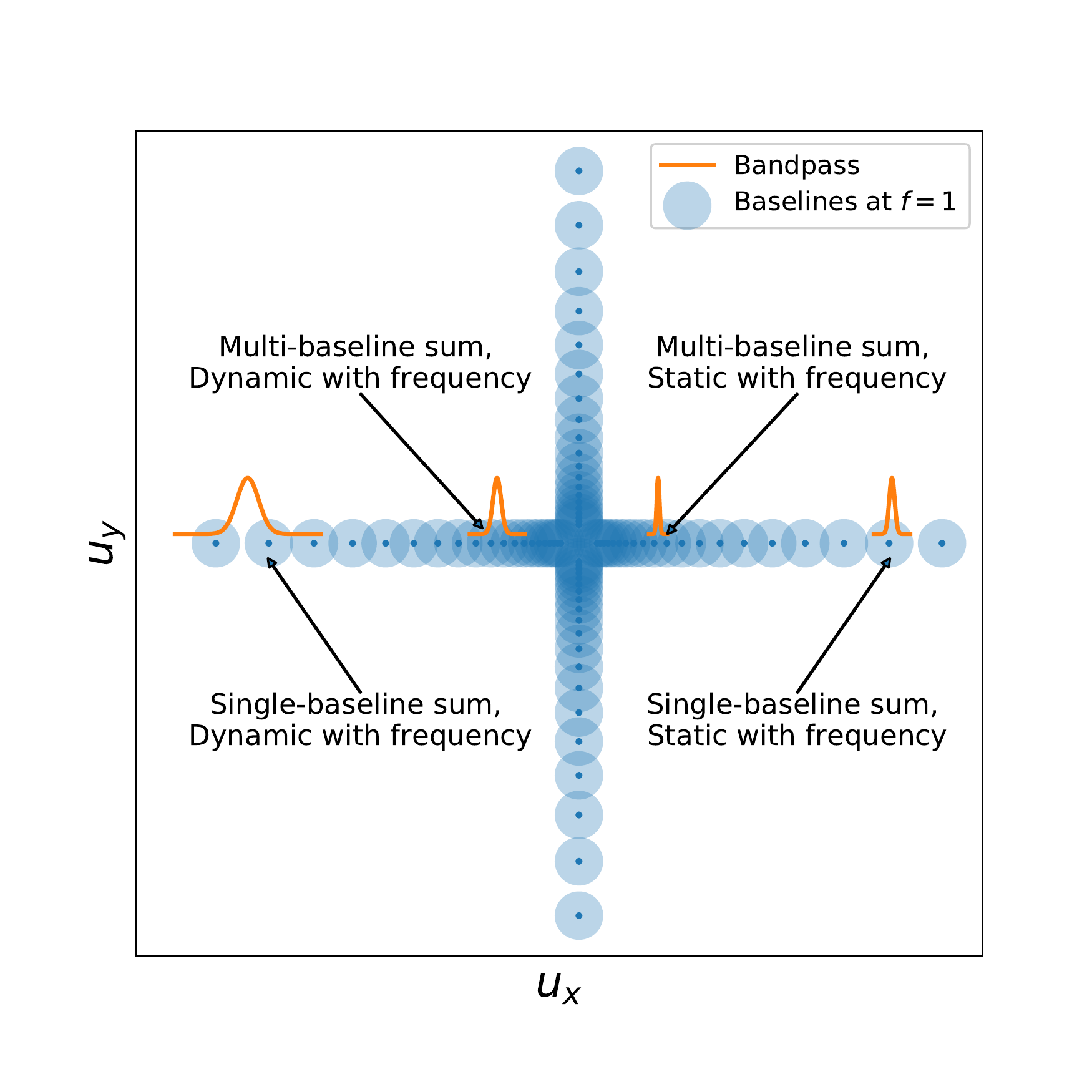}
	\caption{Schematic of weighted gridding, with logarithmic polar grid of baselines. Extent of Fourier-beam kernels are indicated by blue shaded regions around each baseline. Also shown are representative bandpasses, indicating the weight of a baseline centred on the bandpass at $f=1$ as it moves with frequency. The four bandpasses indicate four regimes which may be considered. Note that bandpass indications on the left hand side adopt a lower value of $\tau$. The bandpass (as represented in $u$-space) naturally expands at higher $u$.}
	\label{fig:weighted_gridding_schematic}
\end{figure}

Much can be learned about the effects of including visibility correlations by re-using the logarithmic polar grid baseline layout of \S\ref{sec:classic}, and we address this class of problems in this section.
Here we will dispense with condition (ii) --- that the gridding kernel is arbitrarily narrow --- and use a self-consistent kernel width.
More importantly, we will dispense with the condition that the layout be ``sparse'', 
allowing an arbitrary radial density of baselines (i.e. arbitrarily low values of $\Delta$)\footnote{The adjustment of angular density trivially has no impact, as each ring measures the same expected PS everywhere}. 
It is precisely as we modify this density that we will identify the effects of $uv$-sampling.
We give a schematic of this layout in Fig. \ref{fig:weighted_gridding_schematic}, noting the extent of each baseline via the Fourier beam kernel, and also the interplay of this scale with the bandpass shape.

We note that the limiting case of this derivation -- namely an arbitrarily densely packed set of baselines (i.e. a smooth continuum of baselines) in the radial direction -- is addressed in App. \ref{app:radial}. It is in line with the predictions of the \textit{discrete} polar grid results of this section.

Throughout this section we will use a static beam, and consider the stochastic uniform sky (we have already established that such choices do not greatly impact the qualitative form of the solution). 
Furthermore, due to the fact that the term $|\langle V \rangle|^2$ has negligible power on small scales, we consider only the variance term of Eq.~\ref{eq:power_general}.
Due to the isotropy of the sky and the fact that we use an angularly symmetric layout, we immediately have that $P(\omega, u) = {\rm Var}(\tilde{V}(\omega, u))$ without requiring an integral over $\theta$.
Thus for this section we require only the second term of Eq.~\ref{eq:general_var}, which can alternatively be written:
\begin{subequations}
	\label{eq:wg_master}
	\begin{align}
		{\rm Var}(\tilde{V}) &= \mu_2\nu_0^2 \int d^2\vect{l}\  e^{-l^2/\sigma^2} |I|^2, \\
		I &= \int df\ \frac{\phi_\nu}{W_\nu(\vect{u})}  \sum_{i=1}^{N_{\rm bl}} B(\vect{u} - f\vect{u}_i)e^{-2\pi i f(\omega + \vect{l}\cdot\vect{u}_i)} \nonumber \\
		&= \int df\ \frac{\phi_\nu \sum_{i=1}^{N_{\rm bl}} e^{-2\pi^2\sigma^2(\vect{u}-f\vect{u}_i)^2} e^{-2\pi i f(\omega + \vect{l}\cdot\vect{u}_i)}}{\sum_{i=1}^{N_{\rm bl}} e^{-2\pi^2\sigma^2(\vect{u}-f\vect{u}_i)^2}} 
	\end{align}
\end{subequations}
The solution of $I$ here is in general intractable, due primarily to the sum over baselines in the denominator (i.e. $W_\nu(\vect{u})$).

Eq. \ref{eq:wg_master} may be expanded as follows (with $q^2 = 2\pi^2 \sigma^2$):
\begin{align}
I =  \int df\ \frac{\sum_{i=1}^{N_{\rm bl}} e^{-f^2(\tau^2 + q^2 u_i^2) -2f(\tau^2 + q^2 \vect{u}\cdot\vect{u}_i)-2\pi i f(\omega + \vect{l}\cdot\vect{u}_i)}}{e^{\tau^2} \sum_{j=1}^{N_{\rm bl}} e^{-q^2(f^2 u_j^2 - 2f\vect{u}\cdot\vect{u}_j)}} 
\end{align}
Evaluating this in general is here still intractable.
Nevertheless we can appreciate the general characteristics of the solution by considering the two limits of $u$. 

At small $u$, the exponential cut-off of the beam kernel requires that only baselines with small $u_i$ have non-negligible impact in the sum.
Thus, for $u \ll \tau/q$, for which only terms with $u_i \ll \tau/q$ can contribute, all components with a dependency on $u_i$ in the numerator disappear. 
Furthermore, since the denominator is clearly a much broader function of frequency than the numerator (as $q^2\vect{u}\cdot\vect{u}_i \approx q^2u^2 \ll \tau^2$), it can be removed from the frequency integral. 
Thus we arrive at
\begin{align}
I_{u\ll\tau/q} = \frac{\sum_{i=1}^{N_{\rm bl}}e^{- q^2 u_i^2}}{\sum_{j=1}^{N_{\rm bl}} e^{-q^2(u_j^2 - 2\vect{u}\cdot\vect{u}_j)}} \int df\ \phi_\nu e^{-2\pi i f(\omega + \vect{l}\cdot\vect{u}_i)}. 
\end{align}
The frequency integral is identical to that for the sparse grid (cf. Eq. \ref{eq:stoch_uniform}).
In fact, the solution is a complex sum over terms, all of which are very close to the exact solution of Eq.~\ref{eq:stoch_uniform}, and therefore the behaviour will be almost identical -- i.e. to produce a ``brick'' feature at $u \ll \tau/q$, with an $\omega$ cut-off at $\omega \approx \tau/\sqrt{2}\pi$. 
Remember that this is irrespective of the density of baselines and the width of the gridding kernel, though its regime of applicability is determined by both the gridding kernel width and taper width.

Conversely, at sufficiently large $u$, the baselines are far enough apart from each other that the weighted sum of visibilities is dominated by the single closest baseline (except in rare cases where we consider scales with two or more equidistant baselines, but the rarity of these will make them negligible in the final angular average)
For a given logarithmic separation $\Delta$, this criterion can be considered to be
\begin{equation}
	u^2 \gg \frac{1}{q^2\Delta^2}.
\end{equation}
Note that baselines being separated enough to consider just one in the baseline sum is not equivalent to them being radially separated enough to only consider the same baseline over all frequencies.
It is entirely possible that the baselines will move enough with frequency that they entirely replace one another, while only ever considering one at a time in the baseline sum (cf. Fig. \ref{fig:weighted_gridding_schematic}). 
We denote the baseline closest to $\vect{u}$ at $f$ as $\vect{u}_f$ (this is meant to be an identifier, so that the baseline's value of $\vect{u}$ at $f$ is $f\vect{u}_f$).
In this case, we can rewrite $I$:
\begin{equation}
	\label{eq:Iequation}
	I_{u\gg 1/q^2\Delta^2} = \int df \phi_\nu e^{-2\pi i f (\omega + \vect{l}\cdot \vect{u}_f)}.
\end{equation}

We have already encountered the case in which the baselines are separated enough such that only a single baseline contributes across \textit{all frequencies} (cf. \S\ref{sec:classic}), and this gives the classical form for the wedge. 
This merely shows that if $\Delta$ is large enough, there will \textit{always} be a regime of $u$ such that this classical solution holds for a logarithmic polar grid.

Alternatively, we may consider the limit as $\Delta \rightarrow 0$ (for which the $u$ regime is at extremely large $u$). 
In this case, the closest baseline to $\vect{u}$ will always have $f\vect{u}_f \approx \vect{u}$ (i.e. there will always be a baseline sitting on $\vect{u}$).
Then we have
\begin{equation}
I_{u\gg 1/q^2\Delta^2} = e^{-2\pi i \vect{l}\cdot\vect{u}} \int df \phi_\nu e^{-2\pi i f \omega},
\end{equation}
so that the final power spectrum is
\begin{equation}
	P(\omega, u) = \mu_2 \nu_0^2 \tilde{B}(u)\tilde{\phi}(\omega).
\end{equation}
This separable equation clearly does not contain a wedge, rather containing only the ``brick'' determined by the taper, with an exponential cut-off in $u$.
Though this was shown only for the fictional region $u \rightarrow \infty$ in this case, it is really an example of a continuous distribution of baselines along a radial trajectory, which is shown in detail to omit a wedge in App.~\ref{app:radial}.

Of course, in most cases, the (radial) density of baselines will lie between these extremes, and a natural question is how dense the baselines must be in order to yield a given level of wedge reduction.
We address this question semi-empirically following our conceptual interpretation of the next subsection. 

\subsection{Conceptual Interpretation}
\label{sec:weighted:conceptual}
\begin{figure*}
	\centering
	\includegraphics[width=1\linewidth, trim=0cm 1cm 0cm 0cm]{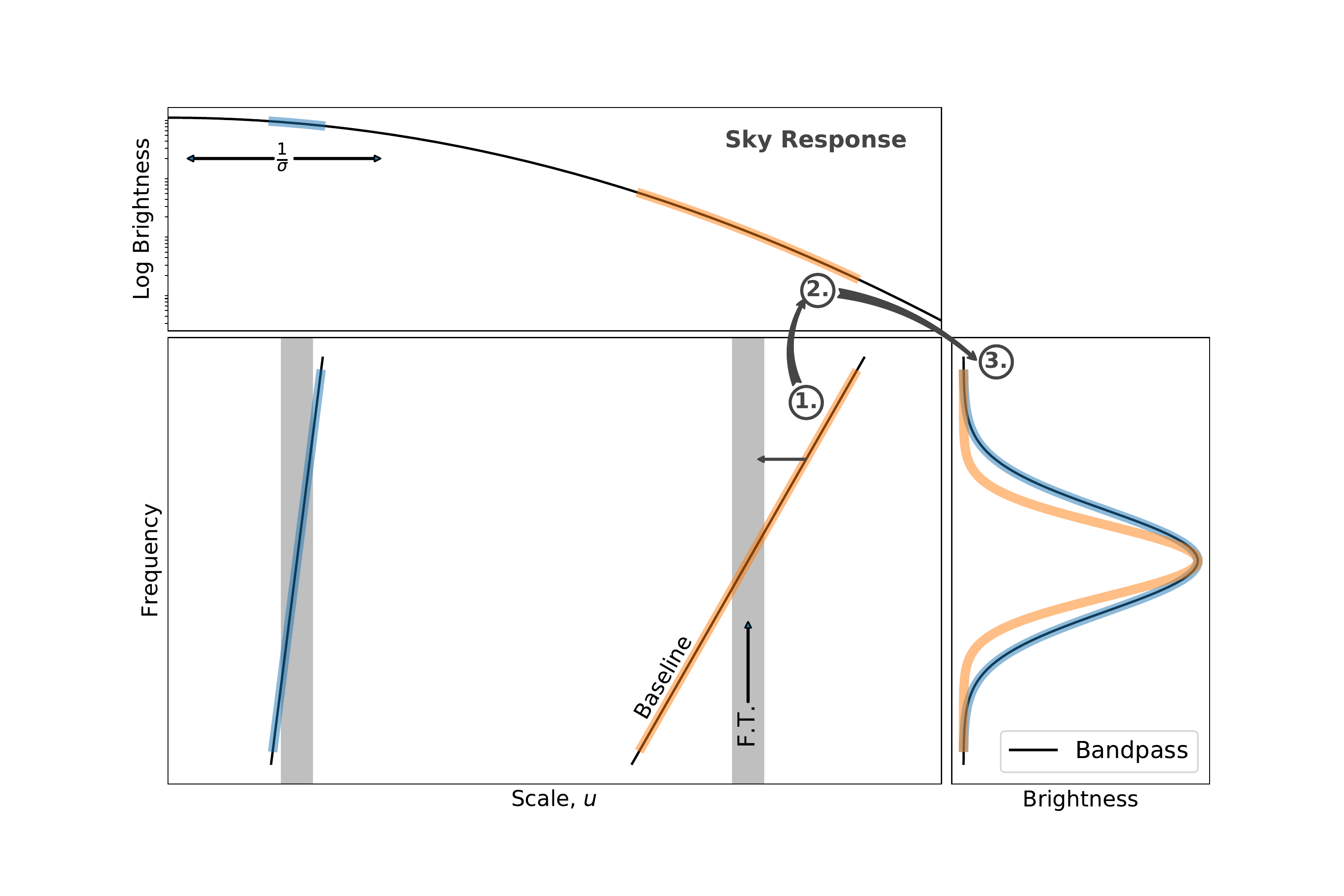}
	\caption{Schematic representation of the mechanics of the emergence of the wedge. Panels should be traced in order of their assigned number. (1.) shows the migration of baselines (black) in $u$ with frequency. Coloured overlaid lines show the ``effective baseline'' traced as an estimate along the points of evaluation, $u$ (grey regions), determined by its weighting kernel. (2.) shows the projection of the effective baseline (coloured) onto the visibility amplitude, accounting for the beam (black line). (3.) shows these results projected (coloured) onto the frequency axis, where they are multiplied by the bandpass (black). The FT of this final curve gives the power spectrum as a function of $\omega$ for the evaluation point $u$. Wider curves transform to narrower curves.}
	\label{fig:wedge_rising}
\end{figure*}

To gain an intuition for the results of the previous subsection, imagine a point $\vect{u}$ for which we are evaluating the power, and consider only baselines that are along a spoke passing through $\vect{u}$ (thus reducing the problem to one dimension). 
Figure~\ref{fig:wedge_rising} gives a schematic representation of this, similar in form to Fig.~\ref{fig:baseline_schematic}.
Here we have chosen a very sparse baseline sampling, akin to the layout chosen for the averaged gridding in the previous section, and show only two points of evaluation (centre of the grey regions), which are concurrent with the baselines at $f=1$.

Performing a FT following the trajectory of a baseline is the delay transform, and at low $u$ this is very close to performing the FT at constant $u$. 
Due to the sparsity of baselines, the effective baseline used in a constant-$u$ transform merely follows the closest baseline.
This ``effective baseline'' is illustrated as a coloured shading overlaying the baseline's trajectory. 
An arrow from this coloured line to the constant-$u$ FT trajectory indicates that it is \textit{this} value of $u$ that is used in the estimation of the Fourier-space visibility.

The top panel shows the immediate results of this effective baseline migration. 
The black curve shows the visibility of the true sky as a function of $u$.
We note that as this is a simplified schematic, we show an effective amplitude of the visibility (which is inherently complex). 
This visibility accounts for the beam attenuation of the instrument, resulting in an exponential curve.
The estimated visibility amplitude at any frequency is merely its value as traced vertically from the coloured curve in the lower left panel. 
That is, in this case, the amplitude is merely traced from left to right as frequency increases, and is indicated by a corresponding coloured line. 
The greater the value of $u$, the larger the arc-length of this line segment, and therefore the greater the ratio between its minimum and maximum. 

The right-hand panel shows these effects on the frequency axis.
In black is the bandpass (or taper). 
To obtain the frequency-space visibility, this is multiplied by the frequency-dependent sky response from the top panel, and shown as corresponding coloured curves. 
While the low-$u$ curve (in blue) is almost constant-amplitude, and therefore barely affects the frequency-space visibility, the high-$u$ curve dramatically suppresses the high-frequency amplitude, effectively causing the frequency-response to be tighter than the natural bandpass (we note that for schematic purposes, we have re-normalised and re-centred the coloured curves).
The frequency-space FT of these curves gives the power spectrum for a given $u$ as a function of $\omega$.
Clearly, tighter curves will transform to wider curves, and hence the ``wedge'' will form when the tightening arising from the effective baseline migration dominates the bandpass (and thence will depend linearly on $u$).
This is much the same description as contained in works such as \citet{Morales2012}, \citet{Parsons2012} and \citet{Trott2012}.

\begin{figure*}
	\centering
	\includegraphics[width=1\linewidth, trim=0cm 1cm 0cm 0cm]{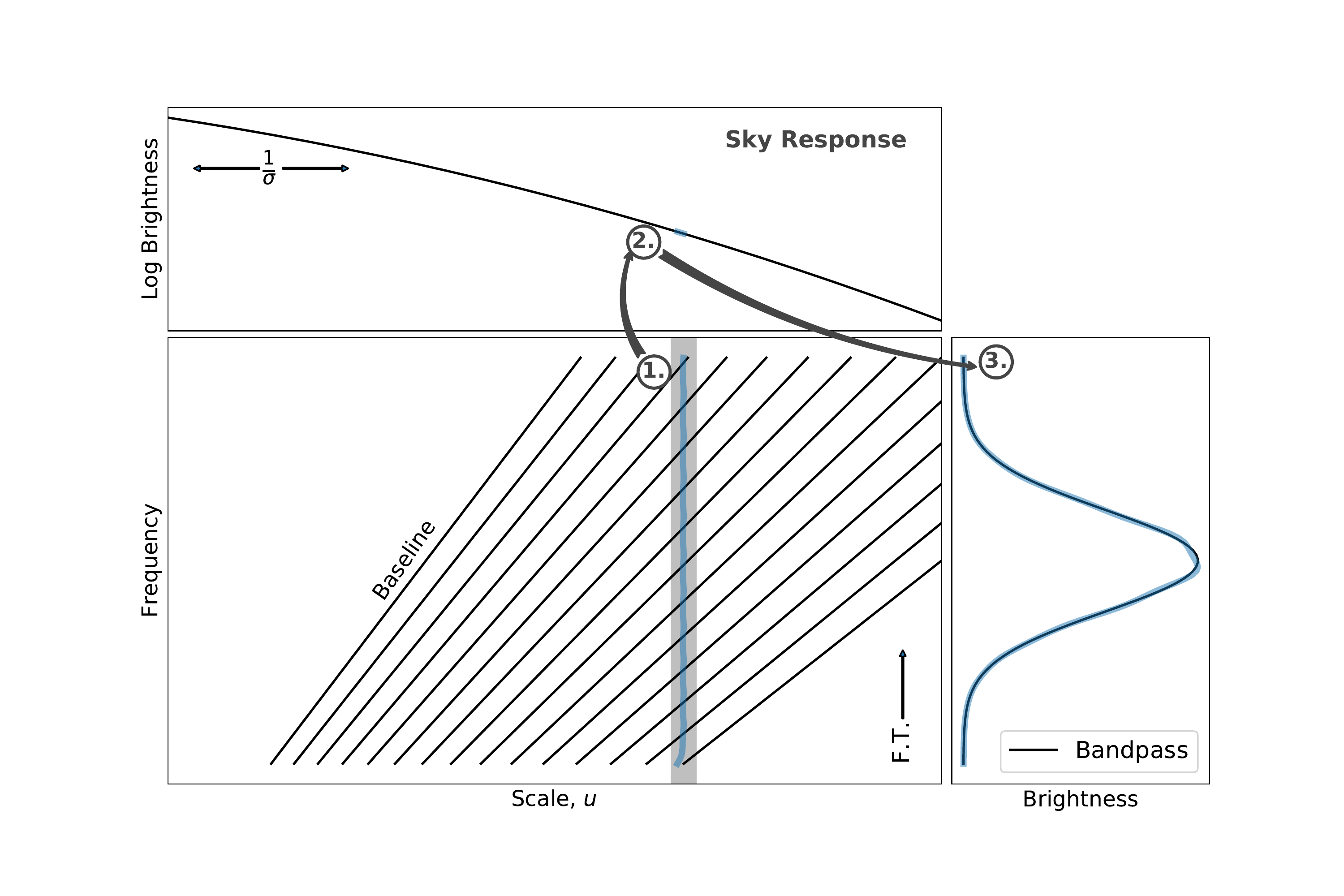}
	\caption{The same as Fig. \ref{fig:wedge_rising}, but for a single evaluation point at high-$u$, with a dense packing of logarithmically-spaced baselines. In this case, the ``effective baseline'' is nearly constant with frequency, and the resulting effect is to negligibly impact the frequency response shape.}
	\label{fig:schematic_inf_bl}
\end{figure*}

Let us consider now a very dense array with logarithmically-spaced baselines.
This we illustrate in Fig. \ref{fig:schematic_inf_bl}.
This figure is the same in form as Fig. \ref{fig:wedge_rising}, but clearly has a much larger number of baselines which pass through the point of evaluation, $u$.
Due to the logarithmic spacing of the baselines, they pass through $u$ at equal intervals of $f$. 
In this case, the ``effective baseline'', shown as the blue curve in the lower-left panel, periodically swaps from one baseline to another. 
We recall that this effective baseline is the weighted average position of all baselines, where the weighting kernel is the Fourier-space beam (in this case, a Gaussian).
Since the baselines are so closely packed, the oscillations created are very small, and it is clear that an infinite number of baselines will yield a truly vertical effective baseline -- corresponding to the true constant-$u$ estimate. 
Consequently, the top-left panel shows that a very small range of visibility amplitudes is covered -- effectively constant over all frequencies.
This in turn renders its product with the bandpass to be solely determined by the latter, and therefore the wedge to be completely avoided. 

\begin{figure*}
	\centering
	\includegraphics[width=1\linewidth, trim=0cm 1cm 0cm 0cm]{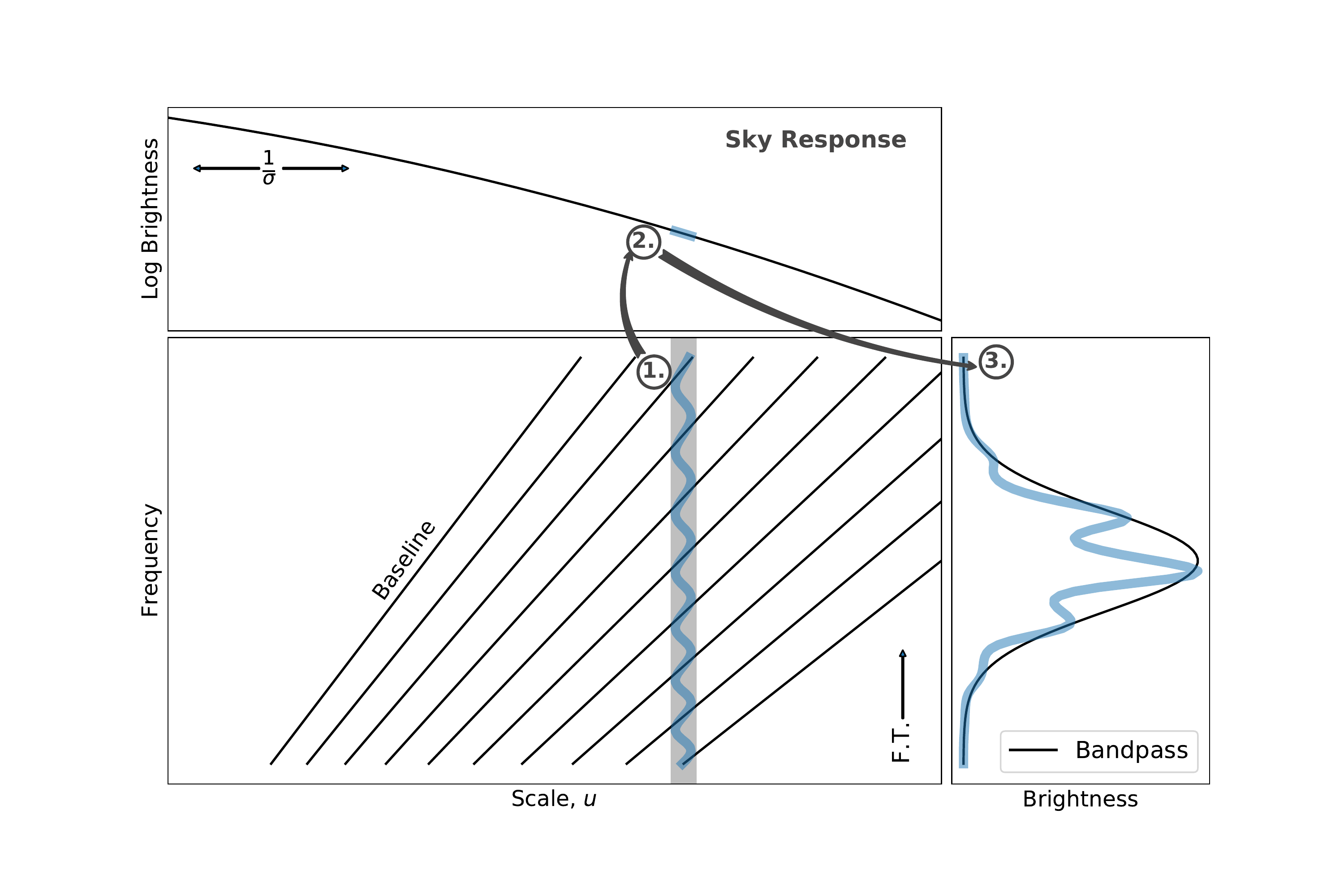}
	\caption{The same as Fig. \ref{fig:wedge_rising}, but for a single evaluation point at high-$u$, with a semi-dense packing of logarithmically-spaced baselines. In this case, the ``effective baseline'' oscillates widely with frequency. Consequently the $u$-footprint is widened and the range of visibility amplitude is also widened. The effect is to overlay a regular oscillatory component atop the bandpass, which contributes a high-$\omega$ ``bar'' (and its harmonics) in addition to the basic ``brick'' yielded by the bandpass.}
	\label{fig:schematic_multi_bl}
\end{figure*}

What of a  baseline density between the previous two figures?
This is shown in Fig. \ref{fig:schematic_multi_bl}.
Here the ``effective baseline'' oscillates between true baselines in a much more marked manner, creating a footprint in $u$ which is much wider than in the previous case.
Projected onto the visibility amplitude, a much wider range is covered, and that range is covered periodically, with period given by the separation of baselines.
Thus it is no surprise to find that the frequency-space product of the response with the bandpass is oscillatory on small scales, with an overall shape given by the bandpass itself. 
The FT of such a function can be approximated as the combination of a smooth Gaussian with width inverse to the width of the bandpass, and a high-frequency term given by the period of the oscillations. 
In this case, in place of a pure wedge, one should obtain a ``bar'' in the 2D PS (along with its harmonics), where the position of the bar in $\omega$-space is inversely proportional to the separation of the baselines, and its amplitude is proportional to this separation.
That is, denser baselines will yield a lower-amplitude bar at higher $\omega$, eventually leading to a negligible bar, and the complete disappearance of the wedge.

In general, baselines will not be regularly logarithmically separated, and various baseline distributions will yield different versions of this oscillatory structure.
For instance, linearly-separated baselines should cause the bar's position to be linearly dependent on $u$.

The exact results of these intuitions are complicated when accounting for baselines in the 2D plane, some of which may significantly contribute to the weight at $\vect{u}$ without being on its radial trajectory.
Furthermore, radial distributions which are not regular will also complicate matters.
We will continue to explore these issues in the following subsections; nevertheless, the basic intuition will remain the same.


\subsection{Wedge Mitigation}

We now turn to considering the properties of wedge mitigation in our simple polar grid baseline layout.
We begin by considering the results of our simple schematic representations of the wedge, in Table \ref{tab:simple_summary}, which suggest that the wedge only emerges from the ``brick'' at $u>\tau/q \equiv \check{u}$.
This defines a region of interest, in which we may hope to mitigate the wedge.
Note that for $u > 1/q\Delta$, the problem can also essentially be considered as 1D, as constructed in \S\ref{sec:weighted:conceptual}, regardless of the angular density.
We shall assume this condition throughout this section, noting that deviations from the assumption will always be small.

Within this region, we ask how a wedge may be \textit{ensured}.
We have already seen that if a single, constant baseline contributes to the sum over baselines, for all frequencies in the bandpass, then we arrive at a wedge.
Thus, we may ensure a wedge by considering the integrated contribution of the two baselines closest to $q$.
If the one baseline dominates, then we are assured of a wedge. 
If the second baseline contributes non-negligibly then we cannot rule out a wedge, but open up the possibility of its mitigation.
We denote the threshold contribution as $e^{-t}$.

In effect, our constraints are defined by the inequality
\begin{equation}
\label{eq:general_wedge_regime}
\frac{\int df \exp\left[-\tau^2 (f-1)^2 - q^2 (u+\Delta_u - fu)^2\right]}{\int df \exp\left[-\tau^2 (f-1)^2 - q^2 (u+\Delta_u)^2 (1 - f)^2\right]} < e^{-t}.
\end{equation}
Here we have assumed that the dominant baseline is co-located with $u$ at $f=1$. 
The qualitative results are insensitive to this assumption.

The solution to Eq. \ref{eq:general_wedge_regime} is 
\begin{equation}
\label{eq:solution_wedge}
\frac{1}{2}\ln\left(\frac{\tau^2 + q^2(u+\Delta_u)^2}{\tau^2 + q^2 u^2}\right) - \frac{\tau^2 q^2\Delta_u^2}{\tau^2+q^2u^2} < -t.
\end{equation}
If we consider only scales where a wedge is possible (i.e. $u>\check{u}$), and maintain that at these scales, $\Delta_u \ll u$, then we may use the approximation $\ln(1+\delta) \approx \delta$ to solve for $\Delta_u$:
\begin{equation}
	\label{eq:solution_for_concentric}
	\Delta_u \gtrsim \frac{u}{2\tau^2}\left(1+\sqrt{1+4\tau^2 t}\right) \approx \frac{u\sqrt{t}}{\tau},
\end{equation} 
where the last approximation assumes $t \gg 1/\tau^2$.
That is, a wedge is \textit{ensured} if the baseline separation is larger than $u\sqrt{t}/\tau$.

The salient features of this equation are 
\begin{enumerate}
	\item  $\Delta_u$ rises proportionally to $u$, so a regular logarithmic spacing for $u > \check{u}$ ensures consistency of wedge/non-wedge for all $u$. 
	\item $\Delta_u$ is inversely proportional to $\tau$, so that larger bandwidths support larger separations before a wedge is ensured.
	\item $\Delta_u$ is proportional to the root of the threshold, $t$. This is difficult to assess conceptually, as we are \textit{a priori} uncertain as to what level the primary baseline must contribute to ensure a wedge.
\end{enumerate}

To get a better sense of the kinds of separations required, we note that 
\begin{equation}
\Delta_u = \Delta_x/\lambda_0 \approx \frac{\Delta_x}{2{\rm m}},
\end{equation}
where $\Delta_x$ is the difference in baseline lengths in distance units (note that this is \textit{not} distances between antennae, but differences between these distances).
Expressing this physical separation in units of the tile diameter, $\Delta_x = \chi D$, we can express our results in terms of the parameter $\chi$.
We note first that
for a (static) Gaussian beam at $\nu_0\approx150$ MHz, the tile diameter can be approximately related to the beam width by
\begin{equation}
 D \approx 1{\rm m}/\sigma,
\end{equation}
Thus, we let $\Delta_u = \chi D/2{\rm m} = \chi/2\sigma$.
In this case, we have that
\begin{equation}
\chi \gtrsim \frac{2\sigma u\sqrt{t}}{\tau}
\end{equation}
ensures a wedge.
As a minimum, at $u= \check{u}$, we have $\chi > \sqrt{t}/\pi$. 

Unfortunately, it is difficult to exactly specify the value of $t$, as it merely represents an order-of-magnitude estimate of the contribution of secondary baselines.
Furthermore, even if we could specify it, we do not have a good analytic handle on what happens for baseline separations smaller than that given by $\chi$ -- we cannot simply assume that the wedge will disappear, though we do expect it to disappear at some small separation.
Thus we turn to a numerical/empirical approach.

\begin{figure*}
	\centering
	\includegraphics[width=\linewidth, trim=1cm 1cm 1cm 0cm]{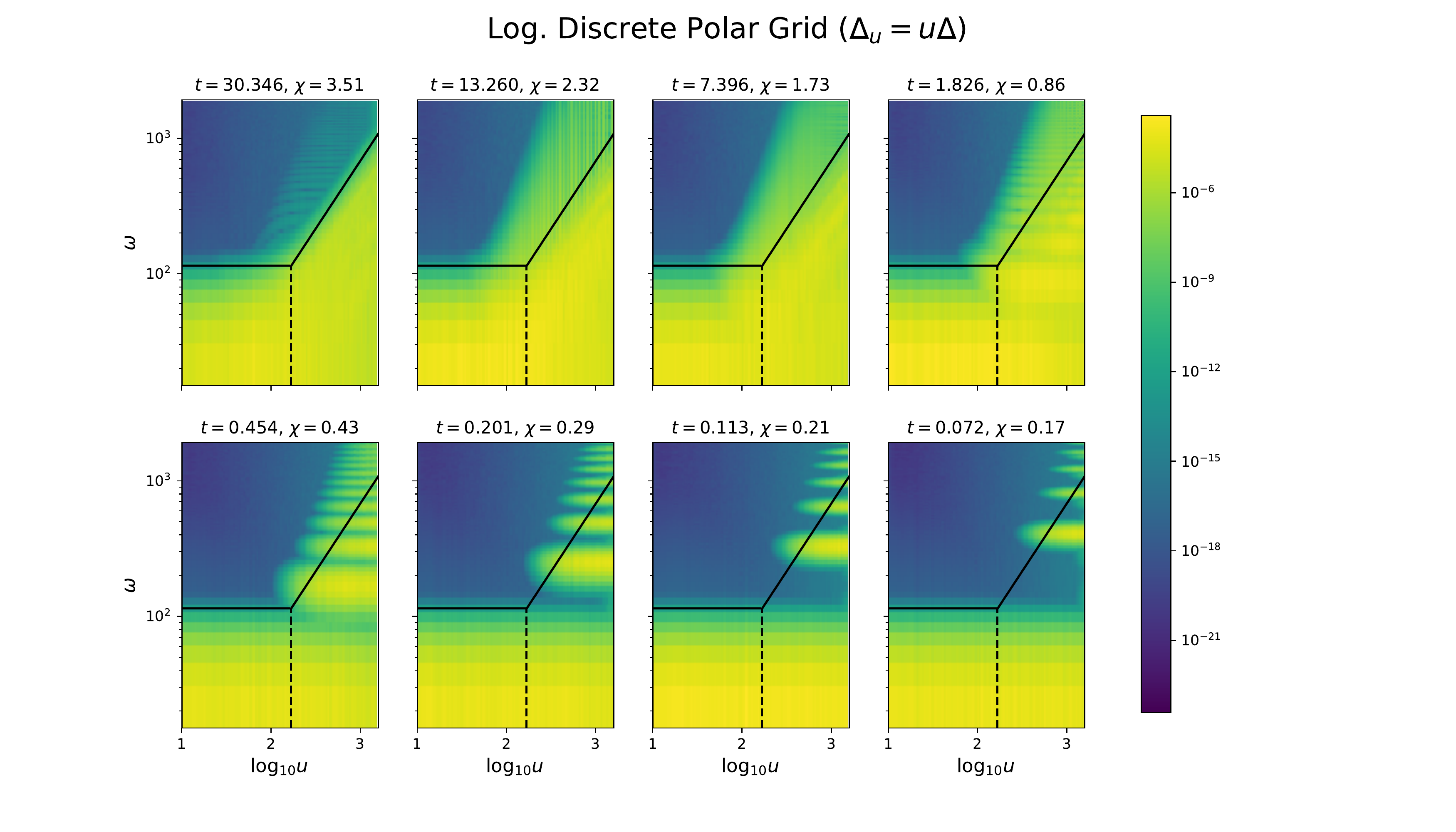} 
	\caption{2D power spectra for fiducial parameters, with static beam and stochastic sky. The baseline layout is concentric, with logarithmically-increasing separations between each circle. The gridding is weighted according to the Fourier beam kernel. Each panel represents a different regular logarithmic spacing, $\Delta$. Titles of each panel indicate the order-of-magnitude contribution of the next-closest baseline at $\check{u}$ (labelled $t$), and the physical separation of baselines as a fraction of the tile diameter at $\check{u}$. The dashed vertical line marks the scale $\check{u}$. The colour-scale in each panel is identical, as are the schematic representations of the wedge/brick shown as black lines (these come from the corresponding row of table \ref{tab:simple_summary}).}
	\label{fig:dense_log_concentric}
\end{figure*}

In Fig. \ref{fig:dense_log_concentric} we show the numerically-calculated power spectra (see App.~\ref{app:numerical} for details on the numerical algorithm) for our fiducial set of physical parameters, and a range of logarithmic separations. 
Each panel is titled by the value of $t$ and $\chi$ which correspond to the baseline separations \textit{at} $\check{u}$ (which is marked by the vertical dashed line).
This clearly shows that a baseline separation of about half the tile diameter is required (taking the minimum, which is at $\check{u}$) for the wedge to begin to disappear.
This occurs at a threshold value of $t \sim 0.4$, corresponding to the second baseline contributing $\sim 60\%$ of the primary baseline over the range of frequencies.
These values are roughly instrument-independent .

Interestingly, we also see horizontal ``bars'' as we had predicted from our conceptual consideration of the problem (cf. \S\ref{sec:weighted:conceptual}).
As predicted, the fundamental bar moves up in $\omega$ as the baselines become closer, and at some separation we expect the harmonics to effectively disappear.

Finally, we ask whether such a layout is physically feasible.
In principle, a perfect polar grid of baselines is unachievable by laying out antennas --- there will always be baselines that are off the grid. 
However, the logarithmic polar grid can be achieved by using logarithmic spokes of antennas, and ignoring all baselines that are off the grid (with a great deal of inefficiency).
In this case, one cannot physically deploy a layout with $\chi < 1$, as the antennas will necessarily overlap with themselves. 
We have found that we require $\chi \approx 1/2$ to mitigate the wedge at $\check{q}$, rendering this completely infeasible. 
While in principle it is possible to design layouts which would enable more closely-spaced baselines, they would come at the cost of reduced layout efficiency, and will be practically infeasible.
We will soon (\S\ref{sec:mitigation:arrays}) explore how leaving the off-grid baselines in the baseline layout affects results.

\subsubsection{Linear Radial Grid}
\label{sec:weighted:linear}
It is interesting to consider the case in which radial baselines are regular in linear space. 
Eq. \ref{eq:solution_for_concentric} suggests that in this case, at some point $u>u'$ the separation will become small enough to mitigate the wedge.
In fact, letting $\Delta_u \equiv \Delta$, we can explicitly solve for $u'$:
\begin{equation}
u' = \tau \Delta/\sqrt{t} \approx 1.5 \tau \Delta,
\end{equation}
where the last approximation assumes $t=0.4$ defines the transition from wedge to no-wedge, as described above. 
Indeed, if $u' < \check{u}$, we expect the wedge to be completely mitigated. 
This is given by the same baseline difference as the logarithmic case, i.e. corresponding to $\chi \approx 1/2$. 
Furthermore, we expect that the bars we saw in the logarithmic case will also be present in the linear case, except that they will not be horizontal, but rather diagonal, as they increase in frequency as $u$ increases.

\begin{figure*}
	\centering
	\includegraphics[width=\linewidth, trim=1cm 1cm 1cm 0cm]{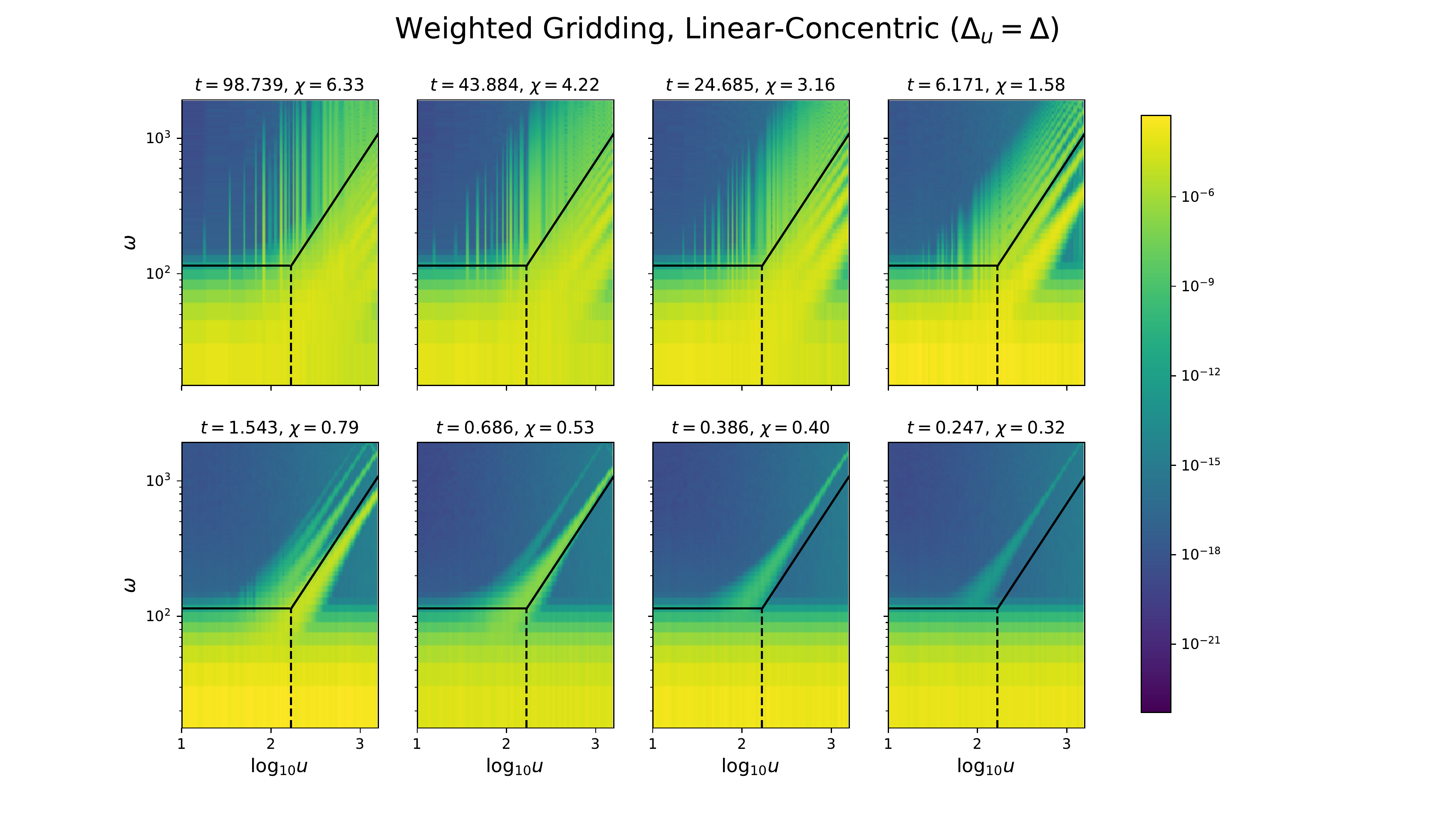} 
	\caption{Exactly the same as Figure \ref{fig:dense_log_concentric}, except that the baselines are spaced regularly in \textit{linear} space.}
	\label{fig:dense_linear_concentric}
\end{figure*}

To illustrate and check these arguments, we show the linear analogue of Fig. \ref{fig:dense_log_concentric} in Fig. \ref{fig:dense_linear_concentric}.
The diagonal bars are quite clear in this case. 
We also see that $t\approx 0.4$ again roughly delineates the disappearance of the wedge at $\check{u}$. 
We note that the vertical lines which appear in the upper panels are due to the fact that in this case, the actual nodes of evaluation, $u_i$, lie at various positions between the radial baselines, rather than being forced to match at $f=1$. 
This creates oscillatory behavior in $u$, but disappears as the distance between baselines increases. 

This kind of layout is \textit{less} physically feasible than the logarithmic polar grid, as it requires impractical baseline densities for all $u$.

\subsubsection{Effects of Angular Density}

Due to isotropy, the results of this section are not sensitive to the number of ``spokes'' in the layout.
Nevertheless it is clear that a single spoke does not have the same \textit{covariance} as a layout with a large number, and when performing inference on a given set of observed data, this covariance is crucial.
An alternative way to think of this is that while a single spoke will yield the same mean over a large random set of skies, it will not necessarily give a good account of a single sky, whose distribution is randomly deviated from perfect symmetry.

\subsubsection{Radial Irregularities}
\label{sec:weighted:irregular}
The precise radial alignment (and regularity) of baselines in the simple polar grid lead to it being impractical layout for wedge mitigation.
In this section, we consider a relaxation of the ideal assumptions of radial regularity in favour of random radial placement, which will come in two forms: (i) completely random and (ii) a random offset from logarithmic regularity.

The advantage of a random array is that it may be perfectly efficient in terms of mapping an antenna layout to the baseline layout --- we need not ignore any pairs within a spoke. 
Thus we can achieve a much greater overall baseline density for the same cost.
Conversely, however, we shall see that the lack of radial alignment increases the overall required baseline density to achieve wedge mitigation.

\begin{figure}
	\centering
	\includegraphics[width=1\linewidth, trim=1cm 0cm 2cm 0cm]{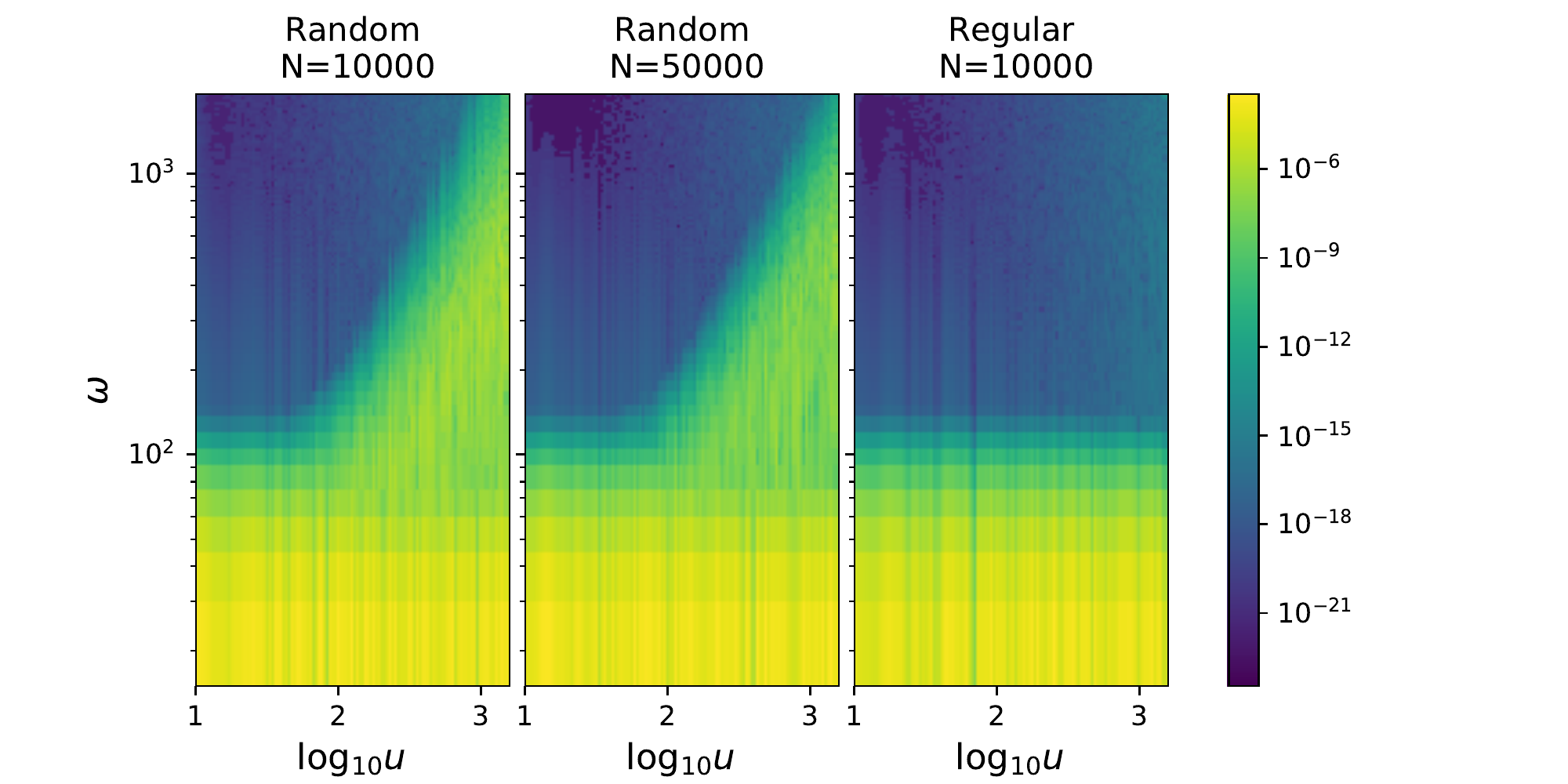}
	\caption{2D PS comparison between baseline layouts for three cases: (i) random radial placement, with mean separation proportional to $u$ and $N=10,000$ (left panel); (ii) the same random placement, but with $N=50,000$ (centre panel), and (iii) regular logarithmic placement of equivalent density to case (i) (right panel). Clearly the introduction of stochastic baseline separations re-introduces a wedge.}
	\label{fig:random_trajectories}
\end{figure}

\begin{figure}
	\centering
	\includegraphics[width=1\linewidth]{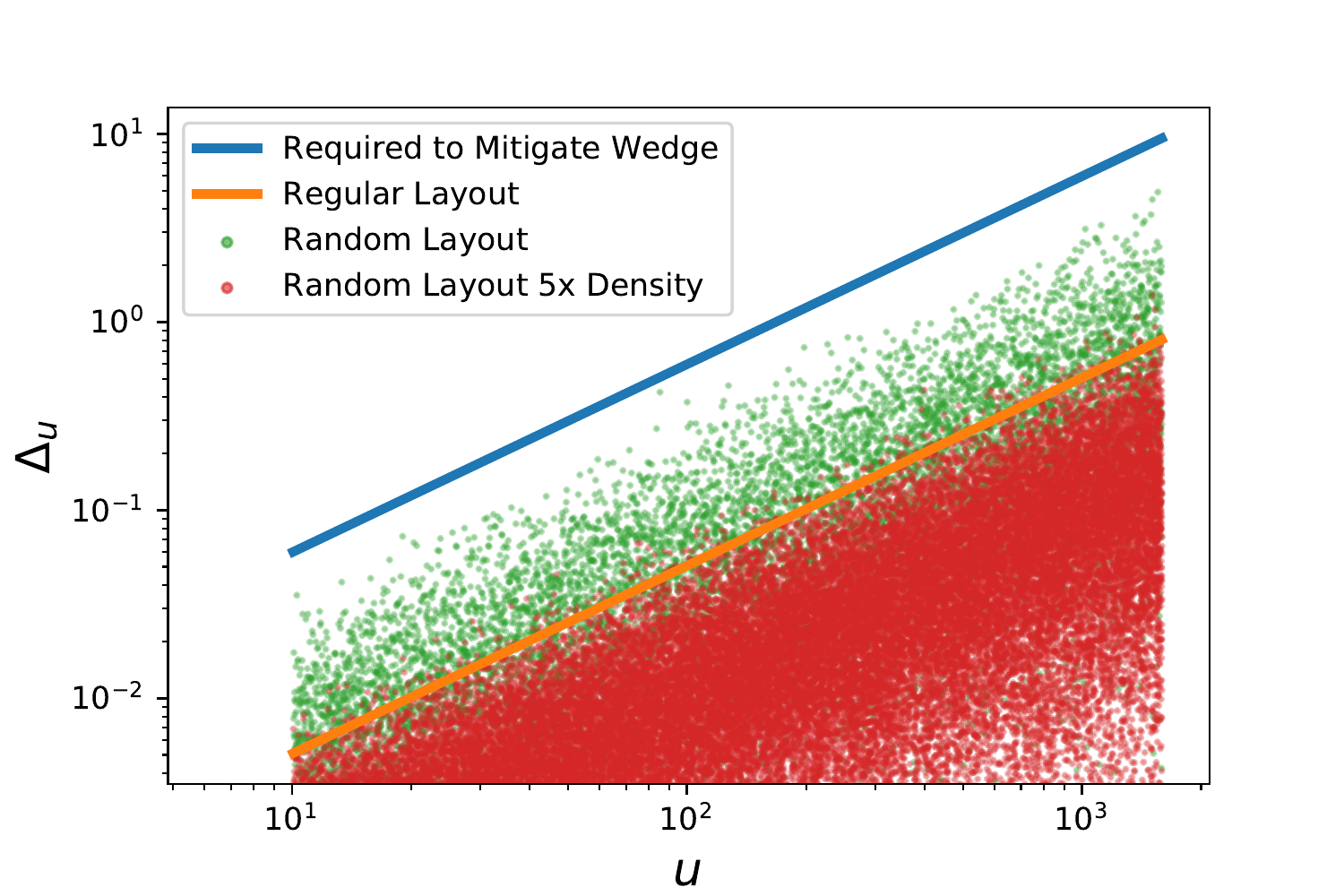}
	\caption{Baseline separations, $\Delta_u$, as a function of $u$, for three cases: (i) random radial placement, with mean separation proportional to $u$ and $N=10,000$ in green; (ii) the same random placement, but with $N=50,000$ in red, and (iii) regular logarithmic placement of equivalent density to case (i) in orange. The blue line shows the threshold separation below which a regular grid significantly mitigates the wedge (cf. Eq. \ref{eq:solution_for_concentric}).}
	\label{fig:random_separations}
\end{figure}

In our ``completely random'' layout, we allow the baselines to be stochastically placed along radial trajectories, with the same overall density as a logarithmic placement. 
We find that doing so yields a somewhat surprising result, which is illustrated in Fig. \ref{fig:random_trajectories}.
In this plot, we compare the 2D PS of a regular logarithmic layout in which the baseline separation is $\Delta_u \approx 0.08\sqrt{0.4}u/\tau$, (i.e. 12.5 times smaller than required to mitigate the wedge), with a layout whose baseline density (and therefore average separation as a function of $u$) is identical, but in which the baselines are stochastically placed. 
We also show the result of an over-dense random layout.
Figure \ref{fig:random_separations} shows the actual separations between baselines as a function of $u$ for each case.
Even for the random arrangements, \textit{all} baselines have separations smaller than the wedge-mitigation threshold.
While we might expect all of them to have near-perfect wedge-mitigation, we find that the random layout yields a subdued, but very present, wedge. 

The explanation for this behaviour arises from the ``bars'' that occur for the regular logarithmic polar grid (cf. Fig. \ref{fig:dense_log_concentric}).
A regular grid creates oscillations which sit atop the bandpass in frequency-space (cf. \S\ref{sec:weighted:conceptual}). 
If the grid is logarithmic in $u$-space, these oscillations are linear in frequency space, causing neatly-spaced peaks in the power spectrum. 
When the baselines have stochastic separations, the oscillations are irregular, and cause a cacophony of ``bars'' above the main ``brick''.
Essentially, this haphazard distribution of peaks restores a somewhat subdued wedge. 

The level to which it is subdued will depend on the baseline density, however it clearly requires a significant increase in density to match the regular logarithmic grid. 
We note that it is not primarily the fluctuating \textit{minimum} separation of baselines that causes the re-emergence of the wedge.
This can be clearly understood from Fig. \ref{fig:random_separations}, in which for the over-dense random layout, the separation very rarely ventures above that of the regular grid. 
The issue is rather that the unevenness of the random distribution causes higher-order structure in the oscillations that lie atop the bandpass, which emanate as the smeared peaks within the wedge.

\begin{figure}
	\centering
	\includegraphics[width=1\linewidth, trim=1.7cm 0cm 3.5cm 0cm]{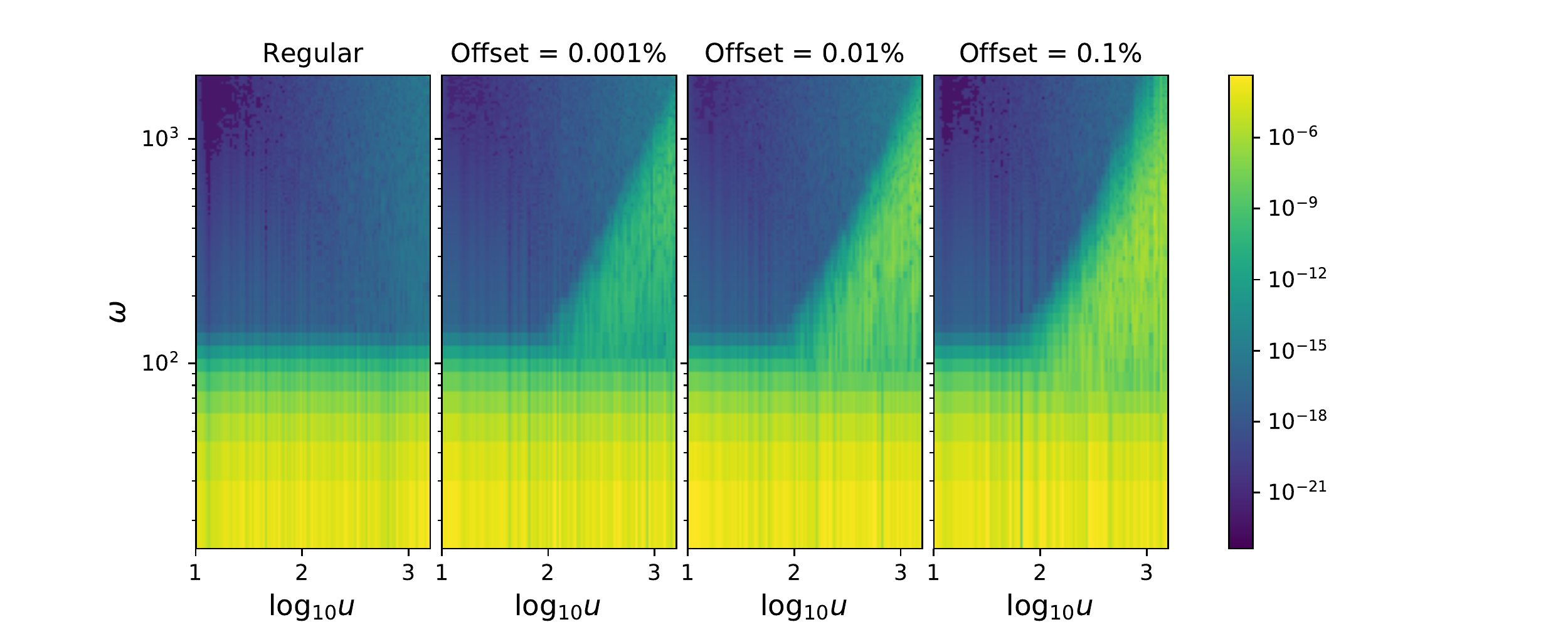}
	\caption{2D PS comparison between baseline layouts for a regular logarithmic polar grid, and three cases in which the baselines have been randomly offset from regularity. A subdued wedge clearly returns even for fractional offsets of $10^{-5}$. The amplitude of the wedge increases by a couple of orders of magnitude as the offsets increase in their magnitude.}
	\label{fig:random_offsets}
\end{figure}

To determine the extent of this effect, we use the same regular set of 10,000 baselines, and randomly offset them by some fraction of their amplitude, according to a normal distribution. 
We show the results in Fig.~\ref{fig:random_offsets}.
Even when the fractional offset is $\sim 10^{-5}$, the wedge returns, albeit quite subdued (2-3 orders of magnitude).
As the offsets increase in magnitude, the wedge is restored to a greater degree, as expected.
It would thus seem that any hopes of mitigating the wedge via regular radial arrays are impractical both due to their high density requirements and their strong dependence on strict regularity. 
Nevertheless, it is possible that irregularity between spokes will alleviate some of this, and we will explore this further in the following section.

\section{Wedge Properties of Real Arrays}
\label{sec:mitigation:arrays}

\begin{table*}[!htb]
	\begin{center}
		\begin{tabular}{ l l l }
			\hline
			\textbf{Label} & \textbf{Description} & \textbf{Varieties}  \\ 
			\hline
			\texttt{circle} & Equi-spaced on circumference of circle, diameter $x_{\rm max}$ &  \\
			\texttt{circle\_filled\_} & Randomly filled circle of diameter $x_{\rm max}$ & Uniform (\texttt{\_0}), Logarithmic (\texttt{\_1}) \\
			\texttt{spokes\_} & Regular radial/angular spacing, max $x_{\rm max}$ & Logarithmic/Linear, $N_{\rm spokes}$  \\
			\texttt{rlx\_boundary} & Equi-spaced on boundary of Reuleaux triangle \citep[eg.][]{Keto1997} &  \\
			\texttt{rlx\_grid\_} & Regular concentric Reulaeux triangles & Logarithmic  \\
			\texttt{hexagon} & Regular hexagon, width $x_{\rm max}$  &  \\
			\hline
			\hline
		\end{tabular}
		\caption{\label{tab:baseline_layouts}Summary of antenna layouts used in Figs. \ref{fig:big_baseline_diagram} -- \ref{fig:big_layout_std}.}
	\end{center}
\end{table*}

We have found that the wedge may in principle be avoided by employing a sufficiently radially dense and regular baseline layout.
In our previous explorations, we have considered explicit baseline layouts, ignoring the fact that no \textit{antenna} layout may exactly correspond to such a baseline layout (alternatively, choosing an antenna layout which corresponds to a superset of the desired baseline layout and ignoring the extraneous baselines).
In this final exploration, we expand our consideration to several physically-feasible antenna layouts, with full correlation of \textit{all} antennas.

In this case, simple ``spoke'' antenna layouts will contain both the regular subset which we have previously considered, and a larger set of irregular baselines. 
Thus we will find whether the increased baseline density outweighs the increased irregularity in terms of wedge mitigation (cf. \S\ref{sec:weighted:irregular}).

\begin{figure*}
	\centering
	\includegraphics[width=1\linewidth, trim=0cm 3cm 0cm 3cm]{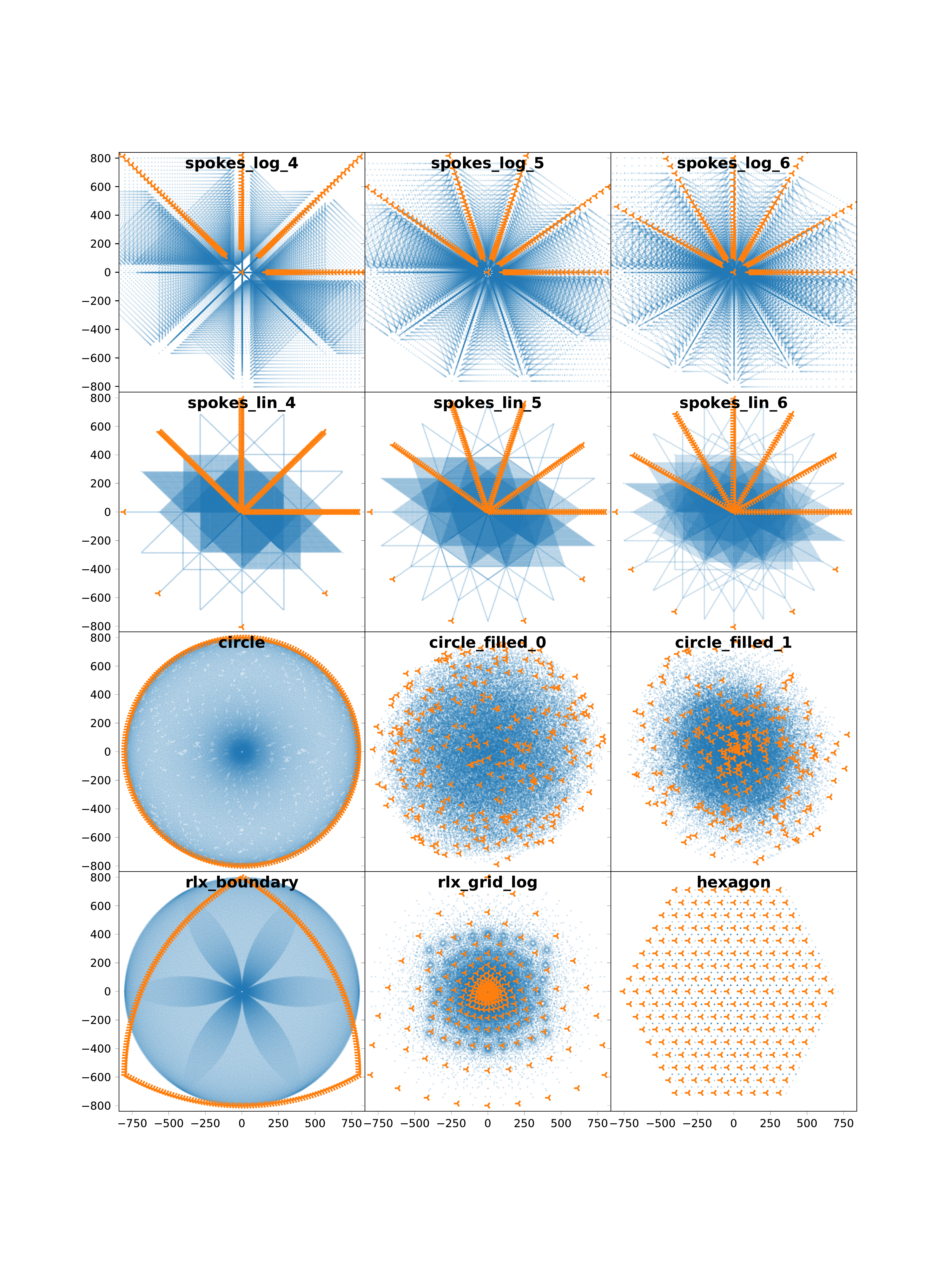}
	\caption{Plots of antenna and baseline layouts for the definitions in Table \ref{tab:baseline_layouts}. The orange markers indicate antennas, and blue points represent baselines. The axes are in units of $u$ at a frequency of 150 MHz. Antennae are all spaced at least 4m apart, which corresponds to the size of an antenna.}
	\label{fig:big_baseline_diagram}
\end{figure*}

The kinds of antenna layouts  we employ (with their variants) can be found in Table \ref{tab:baseline_layouts}, and an illustration of each is found in Fig. \ref{fig:big_baseline_diagram}.
We note that for the linear ``spokes'' layouts, to decrease the redundancy, we use regularly-spaced antennae for half of the spoke, and a single antenna at the far end.
This does not apply for the logarithmic spoke layouts, for which each spoke necessarily begins at the centre.

We use the same number of antennae, $N_{\rm ant}$, for each layout (or as close to this number as possible, given the constraints of some), and place all baselines within a set radius $x_{\rm max} \approx 2u_{\rm max}$.
We choose $N_{\rm ant} = 256$ and $u_{\rm max} = 800$ for the figures in this section.
Each layout is first checked for overlapping antennae, with antenna diameters of 4m (corresponding roughly to the MWA tiles), so that the final layout is physically possible. 
With these choices, comparisons of the power spectra from each array are roughly insensitive to the overall density or ``cost'' of the array, and are rather indicative of the form of the layout itself.
Note that we also use the tile diameter of 4m to calculate the beam width.

Despite these considerations, we must be clear that this is not a test for \textit{how well the layout would recover an underlying 21 cm signal}. 
Specifically, two quantities are of interest for foreground mitigation: the total expected (foreground) power, and its covariance, $\Sigma_P$.  
We have only addressed the expected power in this paper. 
Assuming a reliable model of this quantity can be subtracted from the observed data, the remaining $\Sigma_P$ is the key factor in defining which power spectral modes are usable for averaging to a final one-dimensional power spectrum. 
To a large extent, $\Sigma_P$ is proportional to $P^2$, so the derivations in this paper are indicative of this quantity.
Nevertheless, both sample variance and thermal noise also play an independent role in the determination of $\Sigma_P$, and these are dictated largely by the density of baselines (the former explicitly by the \textit{angular} density).

While the overall density of baselines should be similar in each of the layouts we have chosen -- due to our restriction of setting the antennas within a prescribed radius -- their angular density is decidedly \textit{not}.
We thus expect those with lower angular coverage (eg. the various ``spokes'' layouts) to yield a greater value of $\Sigma_P$, which could hamper 21 cm signal extraction.
We do not pursue a rigorous analysis of these considerations in this paper; our goal is to identify the general effects these layouts have on the establishment of the wedge feature -- not the prospects of 21 cm signal extraction.
Nevertheless, we suggest that such an analysis should be simple enough, by comparing a numerically-generated expectation of $\Sigma_P$ in the presence of both sample variance and thermal noise for each layout.

\ifanalytic
\else
	To determine the expected PS for each layout, we evaluate the PS using the technique outlined in App.~\ref{app:numerical} for the same set of 200 random skies for each layout, taking the mean and standard deviation. 
\fi

\begin{figure*}
	\centering
	\includegraphics[width=1\linewidth, trim=1cm 2cm 0cm 2cm]{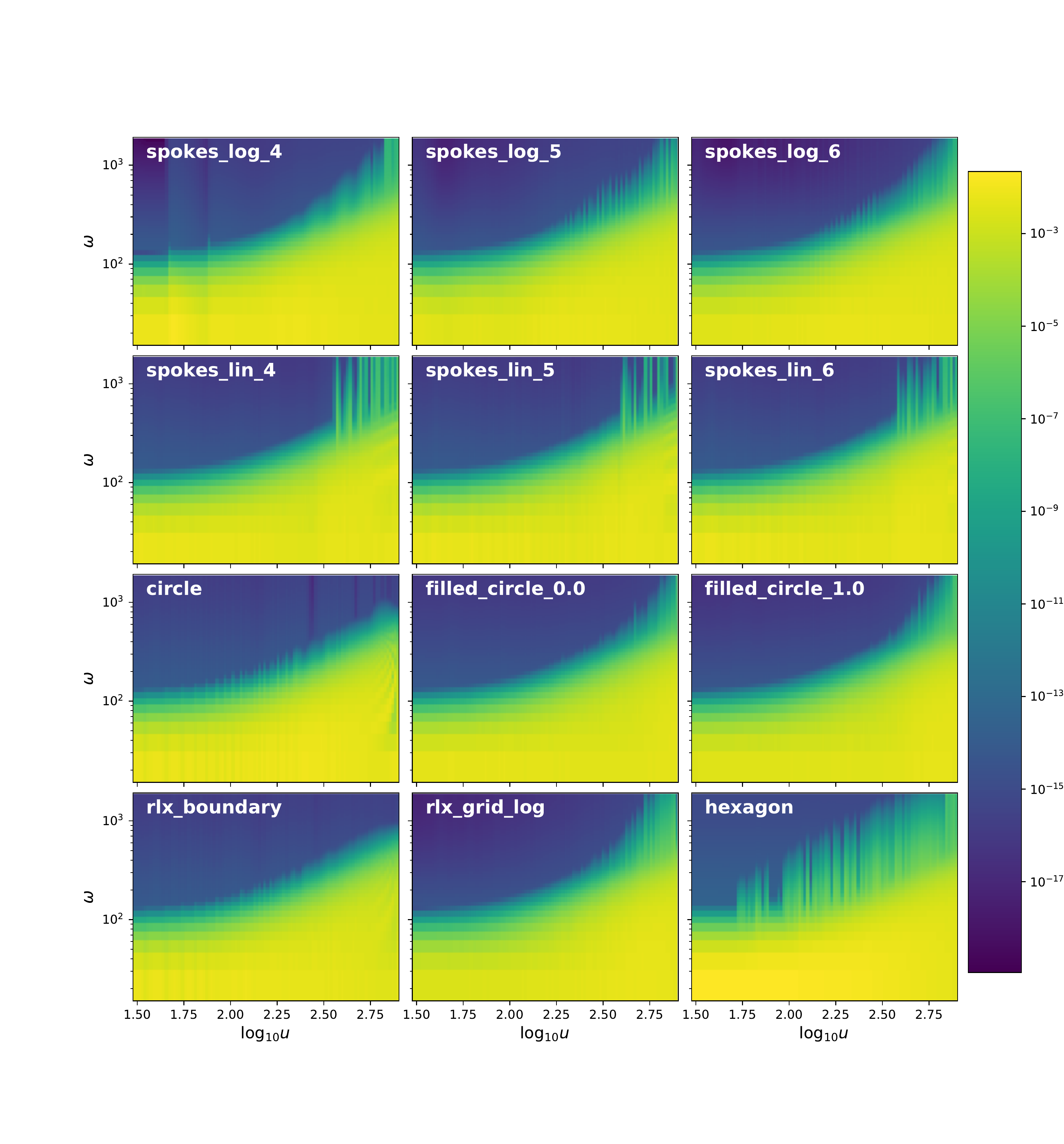}
	\caption{Average 2D PS (over 200 realizations) for each of the layouts we consider (see Table \ref{tab:baseline_layouts} for details of the layouts).}
	\label{fig:big_layout_ps}
\end{figure*}

Figure \ref{fig:big_layout_ps} shows the resulting expected 2D power spectra for each of these layouts.
All layouts show a strong wedge feature with similar shape, as expected, and each exhibits the same bandpass limits (the low-$\omega$ ``brick''). 
Two peculiar features require some explanation.
First, due to our choice of using a gridding kernel which in principle has infinite extent (though in practice, we limit it to 50-$\sigma_u$), grid points $\vect{u}$ which are in extremely sparse UV-sampled locations will tend to evaluate to the same power, as the same distant baseline will be the dominant contributor for all grid points in the region. 
If this is limited to a small arc of the full polar angle, its effect will be negligible, but some of these layouts are extremely sparse for all angles, especially at low-$u$. 
This effect presents as a horizontal `smearing' of the power, and is most noticeable in the low-$u$ modes of the \texttt{spokes\_log\_4} layout. 
A related effect produces the many thin vertical ``spikes'' witnessed at high-$u$ in many of the spectra. 
In this case, however, it seems to be a combination of the irregularity of the baselines with this local sparsity that produces the effect. 
We emphasize that the vertical features, though they appear `noisy', do not disappear as more realizations are averaged, and are therefore systematic. 

\begin{figure*}
	\centering
	\includegraphics[width=1\linewidth, trim=0cm 2cm 0cm 2cm]{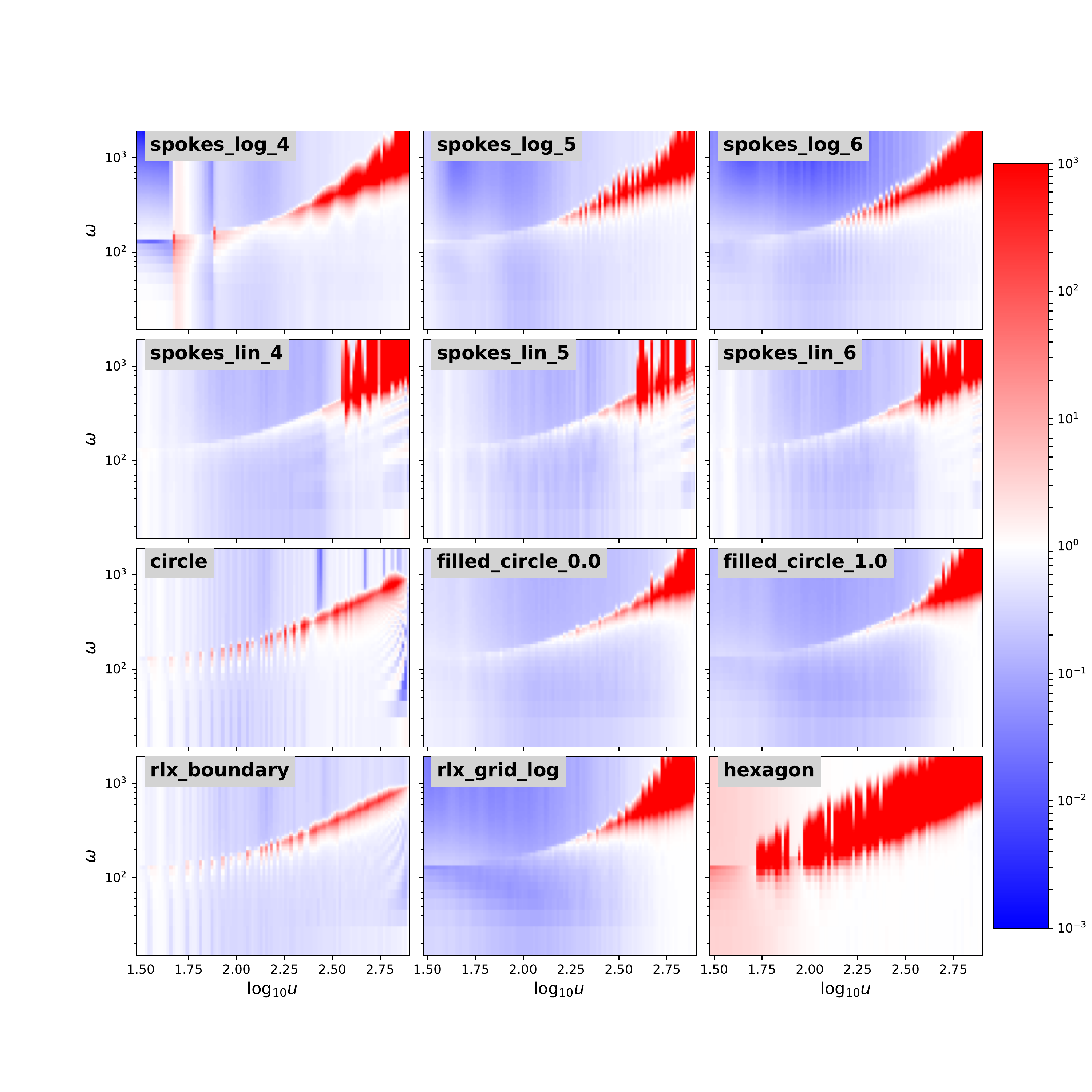}
	\caption{Ratio of 2D PS of each layout in Table \ref{tab:baseline_layouts} to a delay spectrum evaluated with one baseline at each grid point (see \S\ref{sec:mitigation:arrays} for details).}
	\label{fig:big_layout_compare}
\end{figure*}

In Figure \ref{fig:big_layout_compare} we show the ratio of each expected 2D PS against the result of a delay spectrum.
This is precisely the result of \S\ref{sec:classic} (i.e. the limit of sparse baselines), except that each $u$ is assumed to be exactly obtainable.
Thus comparison to this spectrum is appropriate as the sparse limit of baseline density. 
We reiterate that the physical layouts here can be non-local, so that the evaluated power is determined by a relatively distant baseline, whereas the reference `delay spectrum' is always exactly local in this comparison. 

With this in mind, we note that most of the layouts produce less foreground power over most modes than a simple delay spectrum. 
This is to be expected, as the averaging over baselines effectively lowers the amplitude of fluctuations. 
The single exception to this seems to be the hexagonal layout. 
However, on closer inspection, most of the power here is exactly the same as the delay spectrum, as expected from its inherent sparsity. 
At low-$u$, the \texttt{hexagon} is in a region of extreme local sparsity, as discussed above, and therefore cannot be trusted (in the same way as the low-$u$ region of \texttt{spokes\_log\_4}).
The region in and around the wedge does exhibit significantly more power than the delay spectrum, but this is common to all layouts, and we will discuss this momentarily.

The most significant reduction of power occurs in the EoR window for the \texttt{spokes\_log\_6} and \texttt{rlx\_grid\_log} layouts, at 2-3 orders of magnitude.
These layouts have strong logarithmic regularity at the most polar angles compared to other layouts in our sample. 
Though their radial density is not as high as the log-spoke layouts with fewer spokes, it appears that providing some regularity at more angles (and therefore decreasing overall irregularity) outweighs this deficit.
However, these layouts, along with \texttt{circle\_filled\_1}, also have the highest density of short baselines, so it is difficult to isolate the contribution of any single characteristic.

The most visually obvious feature of the ratio plots (figure \ref{fig:big_layout_compare}) is the excess power appearing as irregular vertical stripes protruding from the wedge. 
This power appears to arise due to sparsity of baselines at these scales, such that for a particular grid-point, baselines ``move through'' the grid-point with frequency and leave nulls in the effective spectrum before another baseline passes through. 
This creates a ringing in the Fourier-space spectrum, which throws power outside the wedge. 
This interpretation is supported by the fact that the two layouts which minimize this effect are those with the highest density of baselines at high $u$. 
Conversely, the hexagonal layout, with its extremely sparse and regular baselines, maximizes this effect over much of the range.
This is a well-known key advantage of the delay spectrum, which in principle limits the foreground power exclusively to the theoretical ``horizon line'' for each baseline (notwithstanding other chromatic effects of the instrument and sky).
Nevertheless, it is unclear how this advantage balances against the reduction of window power offered by the dense regular baseline layouts.
Ultimately, these scales, where the density of baselines is low enough to cause this effect, should be ignored in any analysis.

Another interesting feature are the diagonal strips at high $u$ in the \texttt{spokes\_lin} layouts, which appear to be manifestations of the same effect illustrated in Fig. \ref{fig:dense_linear_concentric}, i.e. dense linear radial regularity introducing scale-dependent harmonics in the sky response.
Nevertheless, these are muted compared to the purely regular theoretical arrays previously considered.

\begin{figure*}
	\centering
	\includegraphics[width=1\linewidth, trim=0cm 2cm 0cm 2cm]{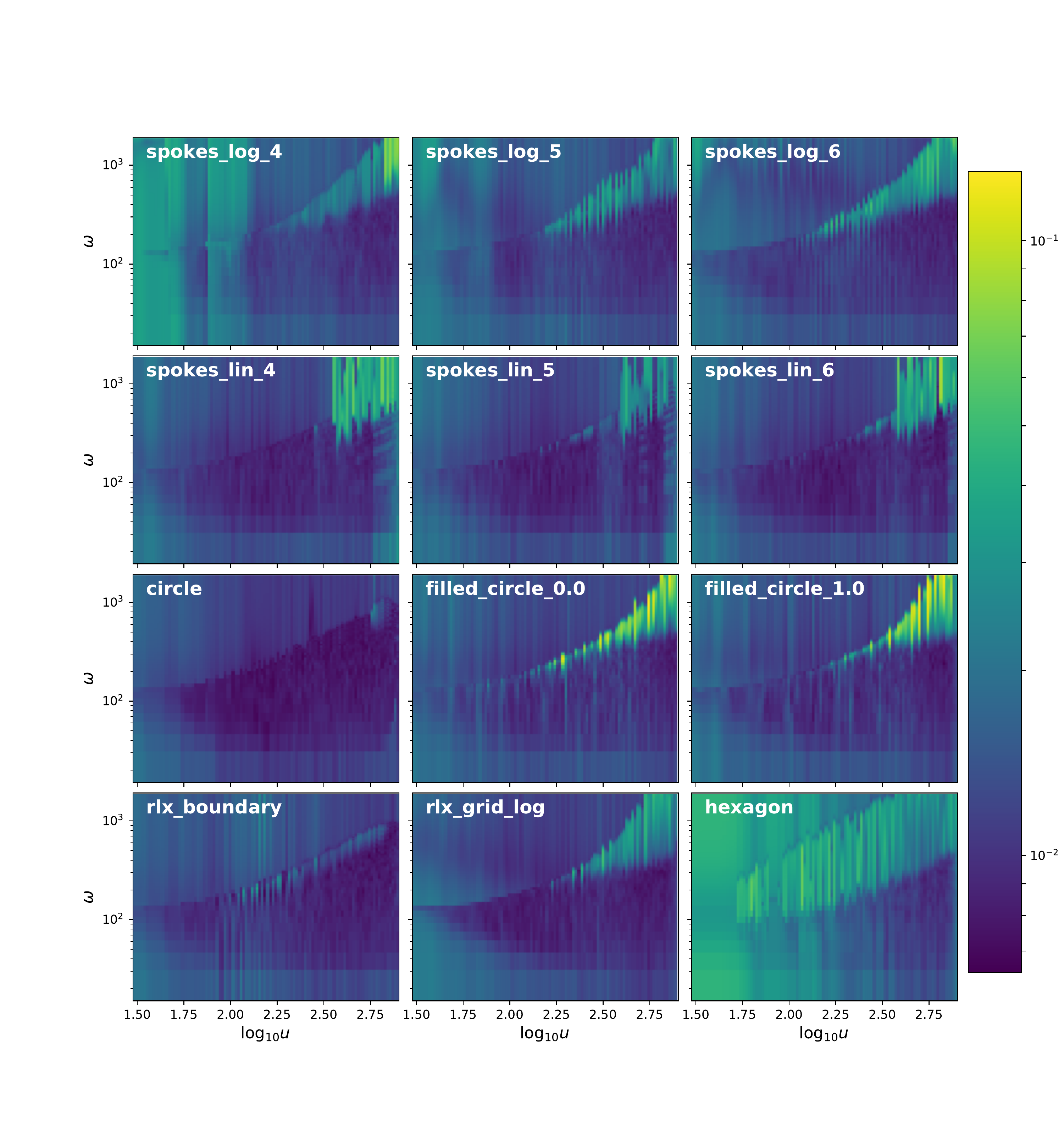}
	\caption{Standard deviation of 2D PS over 200 realizations for each layout in Table \ref{tab:baseline_layouts}.}
	\label{fig:big_layout_std}
\end{figure*}
%

To verify that the results of this section are not subject to high statistical uncertainty, we show the ratio of the standard error of the mean (SEM) to the mean of each PS in Fig. \ref{fig:big_layout_std}.
This illustrates that the mean is accurate to within $\sim10$\%, which means that statistical uncertainty is of minor concern, and that the conceptual results of this section can be trusted in this regard.
Interestingly, the regions of excess power have a relatively high uncertainty compared to the rest of the spectrum, indicating that these regions are more sensitive to the exact positions of point-sources on the sky.
This supports our interpretation that this excess power arises from a dearth of baselines, which would increase the sensitivity of the measured power at a particular $\vect{u}$ to the sky realization, and also increase the variance of measurements over polar angles. 

In summary, with the number of antennae considered, the precise layout has only a minimal effect on the expected PS within the wedge and window. 
Nevertheless, in accord with our semi-analytic considerations of previous sections, it appears that dense logarithmically regular layouts can improve the spectral smoothness of the array, and mitigate foreground power, at the level of 2-3 orders of magnitude. 
We expect this to improve with a higher number of antennae, so that layout considerations will become relatively more important in future high-$N$ arrays.
Conversely, gridding baselines, as opposed to delay transforming on a per-baseline basis, produces artifacts at high-$u$ which can throw excess power out of the wedge. 
It is beyond the scope of this paper to quantitatively assess which method is preferable for measuring the EoR.

\section{Conclusions}
\label{sec:conclusions}

\subsection{Summary}
Using a simple formalism to describe the expected 2D power spectrum of point-source foregrounds (Eqs. \ref{eq:simple_vis}, \ref{eq:vis_gridded}, \ref{eq:vis_full} and \ref{eq:power_general}), we verified the standard schematic `wedge', which has been extensively discussed in the literature.
We then used this formalism, which includes the ability to utilise arbitrary $uv$-sampling functions, to examine the effects of such on the wedge, with the primary conclusion that dense, radially regular layouts can diminish the extent and amplitude of the wedge, but that this effect is small for physically achievable layouts.

Using a semi-analytic approach based on a discrete polar grid $uv$-sampling (largely focusing on logarithmically-separated radial baselines) we find, as suggested in previous works \citep[eg.][]{Bowman2009,Morales2012,Parsons2012}, that increasing the radial density of baselines tends to decrease the amplitude of the wedge (defined as that part of the foreground signature which emerges from the low-$\kpar$ ``brick''). 
Indeed, we find that for regular log-spaced baselines, a density may (in principle) be achieved at which the wedge effectively disappears. 
We explain these results intuitively via radial baseline ``replacement'' with change of frequency (cf. \S\ref{sec:weighted:conceptual}).


Using this semi-analytic model, we explored some of the ramifications of these ideas.
We found that a characteristic separation can be determined which defines the threshold for wedge emergence.
This characteristic separation is proportional to the baseline magnitude $u$ and also to the bandwidth of the observation (cf. Eq. \ref{eq:solution_for_concentric}).
We find that the physical baseline separation (in metres) is approximately $u/\tau$, where our default value for $\tau$ is approximately 100, and in the regime of the wedge, $u > \tau/\sqrt{2}\pi\sigma \approx 150$. 
The minimum baseline separation to mitigate the entire wedge (which occurs at $u=\tau/\sqrt{2}\pi\sigma$, or an antenna separation of $\sim 300$m) is $1/3\sigma \approx 1.7{\rm m}$.
We concluded that such a baseline density is physically impossible for the most efficient antenna layout corresponding to the polar grid baseline layout.
With a less compact array, the baseline density is technically achievable, but highly impractical.

Further, we found that randomising the radial distribution of baselines tends to re-instate the wedge, as the series of overlaid oscillations is smeared out over $\omega \propto \eta$.
Thus the optimal array for wedge mitigation is both dense \textit{and} regular.

Upon examination of the expected 2D PS from some simple antenna layouts, we found that in practice both the window and wedge can be reduced by up to 3 orders of magnitude by employing antenna spokes which are regular in log-space and as dense as possible across many angles.
We noted that such a layout competes with the requirement of angular baseline density to mitigate sample variance.

\subsection{Future Considerations}
The work in this paper paints a rather bleak picture: 
it will be very difficult to combat the wedge via any array design.
Nevertheless, we have shown that in principle, layouts with a higher degree of radial alignment and regularity will serve to reduce the magnitude of the wedge, and therefore potentially yield some increase in the fidelity of future PS estimation.

To establish this rigorously, one needs to consider not only the expected 2D PS, but also its covariance.
These will compete with one another -- the more aligned the layout, the lower the expected wedge, but the higher the overall covariance of the estimate.
A proper analysis of these quantities, and their effect on the signal-to-noise of a fiducial 21 cm signal, is the most pressing future consideration to arise from this work.
Along with this, consideration of non-Gaussian band-pass (or taper) shapes, non-co-planar arrays, non-zenith pointings, and non-flat SED's may be interesting realistic effects to add to the analysis.

\begin{acknowledgements}
	The Centre for All-Sky Astrophysics (CAASTRO) is an Australian Research Council Centre of Excellence, funded by grant CE11E0090. 
	Parts of this research were supported by the Australian Research Council Centre of Excellence for All Sky Astrophysics in 3 Dimensions (ASTRO 3D), through project number CE170100013.
	This research has made use of NASA's Astrophysics Data System.
	All plots in this paper were generated using \textsc{matplotlib}.
\end{acknowledgements}

\clearpage

\begin{appendix}
	
	\section{Derivation of Analytic Examples}
	\label{app:delay}
	
	In this section we derive the solutions to the three analytic examples shown in Table \ref{tab:simple_summary}.
	Recall that in this section, we have employed the following conditions:
	\begin{enumerate}
		\item Sparse discrete polar grid baseline layout
		\item Artificially narrow gridding kernel (width $\rightarrow 0$)
		\item Evaluation at grid points co-located with reference baseline positions.
	\end{enumerate}

	We note that in this scheme, only one baseline may contribute to a given $\vect{u}$ over all frequency. 
	Thus, Eq.~\ref{eq:vis_gridded} is simplified to $V(\nu, \vect{u}) = V_i(\nu,\vect{u}_i)$, removing the sum over baselines.
	
	Finally, since the baselines are arranged symmetrically, the average over $\theta$ is unweighted -- each baseline contributes to the same arc-length.
	The power can thus be simply expressed as
	\begin{equation}
	\label{eq:power_ngp}
	\langle P(\omega, u_\mu) \rangle = \frac{1}{N_\theta} \sum_{k=1}^{N_\theta} 
	{\rm Var}(\tilde{V}(\omega, \vect{u}_{\mu, k})) + \left|\langle \tilde{V}(\omega, \vect{u}_{\mu, k})\rangle \right|^2,
	\end{equation}
	where the sum is over all baselines in a given ring, and the subscript $\mu$ is meant to indicate that we evaluate the visibilities at $\vect{u}_i$, rather than at an arbitrary location.
	Note that evaluation of the PS at $u \neq u_i$ is simple, but will yield the same PS as that evaluated at the closest $u_i$. 
	These solutions therefore really represent a series of step-functions in $u$, where each step is centred on $u_i$.

	\subsection{Sky Model: Single Source}
	Within this subsection we will consider a single-source sky at $\vect{l} = \vect{l}_0$ with $S=S_0$.
	Due to the non-stochastic nature of the sky we only require the mean visibility, which is simply:
	
	%
	%
	\begin{align}
	\langle \tilde{V}(\omega, \vect{u}_\mu) \rangle = \nu_0 S_0 \int df\ \phi_\nu  B_\nu(l_0) e^{-2\pi i f(\omega + \vect{u}_\mu \cdot \vect{l}_0)}.
	\end{align}
	
	
	\subsubsection{Static Beam} 
	If the beam is frequency-independent, it comes out of the integral and we are left with
	\begin{equation}
	\label{eq:tmp0}
	\tilde{V}(\omega, \mathbf{u}_\mu) = S_0 \nu_0 B(l_0) \int df e^{-\tau^2 (f-1)^2} e^{-2\pi if (\omega +\vect{u}_\mu \cdot\vect{l}_0)}.
	\end{equation}
	
	Here we make use of the following useful identity, and we shall repeatedly do so throughout this section:
	\begin{equation}
	\label{eq:int_of_exp}
	\int_{-\infty}^{+\infty} e^{-ax^2 - bx + c}dx = \sqrt{\frac{\pi}{a}}e^{b^2/4a + c}.
	\end{equation}
	Rearranging Eq. \ref{eq:tmp0}, we find 
	\begin{align}
	a &= \tau^2,\nonumber \\
	b &= 2\pi i (\omega + \vect{u}\cdot \vect{l}_0) - 2  \tau^2, \nonumber \\
	c &= -\frac{l_0^2}{2\sigma^2} - \tau^2.
	\end{align}
	This yields
	\begin{align}
	\label{eq:vis_ss_sb}
	\tilde{V}(\omega, \mathbf{u}_\mu) =  &\frac{S_0 \nu_0 \sqrt{\pi}}{\tau} \exp\left(-\frac{l_0^2}{2\sigma^2}\right) \nonumber \\
	& \times\exp\left( -\frac{\pi^2(\omega + \vect{u}_\mu\cdot \vect{l}_0)^2}{\tau^2}  - 2 i\pi (\omega + \vect{u}_\mu\cdot \vect{l}_0)\right)
	\end{align}
	
	Furthermore, the power can be written:
	\begin{align}
	\label{eq:ss_ngp_static}
	P(\omega, u_\mu) = \frac{1}{N_\theta}  & \frac{S_0^2 \nu_0^2 \pi}{\tau^2} \exp\left(-\frac{l_0^2}{\sigma^2}\right) \nonumber \\
	& \times \sum_{k=1}^{N_\theta} \exp\left( -\frac{2\pi^2(\omega + u_\mu l_0 \cos (2\pi k/N_\theta))^2}{\tau^2} \right).
	\end{align}
	This sum has no general closed form solution.
	Nevertheless, it is not difficult to ascertain its general behaviour. 
	The two terms in the exponential will compete for dominance, and since cosine has a maximum of unity, we can determine a line of equal weight: $\omega = ul_0$.
	When the $\omega$ term is dominant, the integrand loses sensitivity to $\theta$, and the power can be written
	\begin{equation}
	P(\omega \gg ul_0, u) = \frac{S_0^2 \nu_0^2 \pi}{\tau^2} \exp\left(-\frac{l_0^2}{\sigma^2}\right)  \exp\left( -\frac{2\pi^2 \omega^2}{\tau^2} \right),
	\end{equation}
	Thus we expect that there will be a (sharp) exponential drop in the power for $\omega \gg ul_0$. 
	When $ul_0$ is small, the entire function $P(\omega)$ (i.e. a vertical line in the 2D PS) will obey this equation, and the cutoff will appear at a characteristic scale $\omega \sim \tau/\pi\sqrt{2}$. 
	Larger $ul_0$ acts as a buffer, requiring $\omega$ to overcome it before the exponential drop is realised (at a much sharper rate, due to the increased amplitude of the exponential). 
	The exact point at which $\omega$ overcomes the $ul_0$ term is difficult to obtain in closed form (it can easily be obtained as a power series), but we merely state the empirical result that it is close to $ul_0$. 
	Thus we have a cutoff at $\omega \approx {\rm max}(\tau/\pi\sqrt{2}, ul_0)$, where the first limit defines a ``brick'' at low $(\omega,u)$, and the second defines a ``wedge'' at higher $u$.

	\subsubsection{Chromatic Beam}
	\citet{Liu2014} have pointed out that regardless of whether the beam is chromatic or not, it is a much broader function of frequency than the taper, and therefore it is a good approximation to bring it outside the integral, and evaluate it at $f=1$. 
	This would yield precisely the same result as the achromatic beam of the previous section.
	Nevertheless, we wish to present an exact formula in this section.
	
	In this case, the only aspect that changes from the previous section is that we have $a \rightarrow \tau^2 + l_0^2/2\sigma^2$ since the beam moves back inside the integral.
	Thus we achieve
	\begin{align}
	P(\omega, u_\mu) =  & \frac{S_0^2 \nu_0^2 \pi}{N_\theta p^2} \exp\left[-\tau^2 \left(1 - \frac{\tau^2}{p^2}\right)\right] \nonumber \\
	& \times \sum_{k=1}^{N_\theta} \exp\left( -\frac{2\pi^2 (\omega + u l_0 \cos (2\pi k/N_\theta))^2}{p^2}\right),
	\end{align}
	with $p^2 = \tau^2+l_0^2/2\sigma^2$.
	The behaviour of this equation is very similar to the previous static case, except that the effect of $\tau$ is balanced by the effect of the beam-width. 
	That is, setting $\tau$ arbitrarily small (i.e. very wide band-pass) will no longer yield an arbitrarily tight ``brick'', as the beam-width will have the effect of broadening it.
	
	In practice, for instruments targeted at observing the EoR, $\tau^2 \gg 1/2\sigma^2$, so that the achromatic beam  is a reasonable approximation, as expected by the arguments from \citet{Liu2014}.
	
	\subsection{Sky Model: Stochastic Uniform}
		
	We merely need to solve Eqs. \ref{eq:su_meanvis} and \ref{eq:su_covvis} for a Gaussian beam, and then integrate over frequency.
	We evaluate only for a static beam in this case, as we have already seen that a chromatic beam is too broad (in frequency) to have a significant impact on the result.
	
	The mean term is simply
	\begin{align}
		\langle V(\omega, u_\mu) \rangle &= 2\pi \sigma^2 \nu_0\bar{S} \int df\ e^{-2\pi i f\omega} \phi_\nu  e^{-2\pi^2 \sigma_\nu^2 f^2 u^2}.
	\end{align}

	Using the identity Eq.~\ref{eq:int_of_exp}, and noting that the equation depends only on $u^2$ and therefore needs not be integrated around the annulus, we find
	\begin{align}
	|\langle \tilde{V}\rangle|^2 &= \frac{4 \bar{S}^2 \nu_0^2 \pi^3 \sigma^4}{\tau^2 + 2\pi^2 \sigma^2 u^2} \exp\left[ 2\left(\frac{\tau^4 - \pi^2\omega^2}{\tau^2 + 2\pi^2\sigma^2 u^2} - \tau^2 \right) \right].
	\end{align}
	
	To begin the variance, we use Eq.~\ref{eq:su_covvis}, along with the various assumptions we have thus far made, to obtain
	\begin{align}
		\label{eq:stoch_uniform}
		{\rm Var}(\tilde{V}) = \mu_2 \nu_0^2 & \int d\vect{l}e^{-l^2/\sigma^2} \left| \int df   e^{-2\pi if(\omega + \vect{u}\cdot \vect{l})} \phi_\nu  \right|^2
	\end{align}
	We use the result of (Eq. \ref{eq:vis_ss_sb}) directly to obtain
	\begin{align}
		{\rm Var}(\tilde{V}) = \frac{\pi \mu_2 \nu_0^2}{\tau^2} \int d\vect{l}\ \  e^{-l^2/\sigma^2} 
		\exp\left(-\frac{2\pi^2(\omega+\vect{u}_\mu\cdot\vect{l})^2}{\tau^2} \right) .
	\end{align}
	Due to statistical isotropy, we may without loss of generality evaluate the case $\vect{u} = (u,0)$, and perform the integral over $\vect{l}$ in 2D Cartesian space, to finally find
	\begin{align}
	\label{eq:sparse_solution}
	{\rm Var}(\tilde{V}) = \frac{\mu_2 \nu_0^2 \pi^3 \sigma^2}{\tau\sqrt{\tau^2 + 2\pi^2\sigma^2u^2}} \exp\left(-\frac{2\pi^2 \omega^2}{2\pi^2\sigma^2u^2+\tau^2}\right).
	\end{align}
	
	Now combining both terms of the power spectrum, we can make some simple observations.
	Firstly, for $\pi u \sigma \ll \tau$, we have
	\begin{equation}
	P_{\pi u \sigma \ll \tau} = \frac{\nu_0^2 \pi^2 \sigma^2}{\tau^2} e^{-2\pi^2 \omega^2/\tau^2} \left(\bar{S}^2 \sigma^2 + \mu_2 \pi \right).
	\end{equation}
	This has a sharp cut-off at $\omega \approx \tau/\sqrt{2}\pi$, creating the familiar lower-left ``brick'' in the 2D PS.
	Conversely, we have
	\begin{align}
	P_{\pi u \sigma \gg \tau} &= \frac{\nu_0^2 \sigma}{u} e^{-\omega^2/2u^2\sigma^2}\left[\frac{ 2\pi  \sigma \bar{S}^2 e^{-\tau^2}}{ u}  + \frac{\mu_2 \pi e^{-\omega^2/2\sigma^2 u^2}}{\sqrt{2}\tau}\right] \nonumber \\
	&\approx \frac{\nu_0^2\mu_2 \pi \sigma}{\sqrt{2}\tau u} e^{-\omega^2/u^2\sigma^2}, 
	\end{align}
	where the final line assumes that $\omega < \tau^2$, which covers all the reasonable values of $\omega$.

	This clearly has a sharp cut-off at $\omega = u\sigma$, creating the wedge (cf. rightmost panel of Fig. \ref{fig:ss_ngp_static}).

\section{Radially Smooth Layout}
\label{app:radial}
Here we consider a polar grid layout in which the radial spokes are no longer discrete but are of such high density that they may be considered smooth.
This will allow us to derive some constraints on how `smooth' the radial distribution of baselines must be to avoid a wedge.

Let $\rho = \rho_\theta \rho_u$ be the density of baselines as a function of $u$ and $\theta$.
Then the sums over baselines in Eq. \ref{eq:wg_master} reduce to integrals over $\rho$: 
\begin{align}
I = \int df \frac{\phi_\nu}{W_\nu} \int d^2 \vect{u}_i \frac{\rho_\theta \rho_u}{u_i} e^{-q^2(\vect{u} - f\vect{u}_i)^2} e^{-2if(\omega +\vect{l}\cdot\vect{u}_i)}.
\end{align}

We may calculate the total weight, performing the integration in polar co-ordinates, making the substitution $u'_i = fu_i$:
\begin{align}
W_\nu = &\frac{e^{-2\pi^2 \sigma^2u^2}}{2\pi f^2} \int_0^{2\pi} d\theta\  \rho_\theta \nonumber \\
&\times \int du'_i\  \rho_u(u'_i/f^2) e^{-q^2({u'_j}^2 - 2uu'_i\cos\theta )}.
\end{align}
It is difficult to proceed further without specifying some form for $\rho_u$. 
Nevertheless, we note that $\rho_u$ will only contribute to the $u'_i$ integral if it is sufficiently sharply peaked -- otherwise it can be treated as a constant and removed from the integral.
We let $\rho_u$ be an arbitrary linear combination of Gaussians, centered around points $u_l$:
\begin{equation}
\rho_u \propto \sum_l a_l \exp\left(-\frac{(u'_j - u_l)^2}{2 \sigma_l^2}\right),
\end{equation}
where the normalisation constant is irrelevant as it cancels in the final visibility. 

The equation for $W_\nu$ may thus be re-written as
\begin{align}
W^T_j = &\frac{e^{-q^2 u^2}}{f^2} \int d\theta\  \rho_\theta \sum_l a_l e^\frac{-u_l^2}{2f^4\sigma_l^2} \nonumber \\
&\times \int du'_i\  \exp\left(-{u'}_i^2(q^2 + \frac{1}{2\sigma_l^2})\right. \nonumber \\
&+ \left. 2u'_i(q^2 u\cos \theta + \frac{u_l}{2f^2\sigma_l^2})\right).
\end{align}
Performing the $u'_i$ integral, each term in the sum becomes
\begin{equation}
a_l \sqrt{\frac{\pi}{q^2 + \frac{1}{2\sigma_l^2}}} \exp\left(\frac{q^2\left[q^2 u^2\cos\theta/2 + \frac{uu_l\cos\theta}{2f^2\sigma_l^2} - \frac{u_l^2}{f^4 \sigma_l^2}\right]}{2q^2 + \frac{1}{\sigma_l^2}}\right).
\end{equation}
If $\sigma_l \gg 1/2q = 1/2\pi\sigma$, then we can ignore the $\sigma_l$ term in both the square root and the denominator of the exponential. 
In fact, this same inequality also reduces the numerator to its first term (for $f\sim 1$), such that the form for $W_\nu$ is
\begin{align}
W_\nu = &\sqrt{\frac{\pi}{q^2}} \frac{e^{-q^2u^2}}{f^2}   \sum_l a_l \int d\theta\ \rho_\theta \exp\left(\frac{q^2 u^2\cos\theta}{4}\right).
\end{align}

The condition that $\sigma_l \gg 1/2\pi\sigma$ for all terms $l$ is thus a well-specified ``smoothness" bound, though we note that it is a conservative bound;
even if a term is \textit{more} peaked than permitted by the bound, if it has a small relative amplitude then its contribution may be ignored. 
This is important for real arrays, in which the baselines form delta-functions in the UV plane. 
Though every point consists of a ``Gaussian'' which is more peaked than the bound, they may be spaced closely enough that each of them contributes negligible weight, thereby approximating a ``smooth" array.

This smoothness bound, for a realistic array at $\nu_0 \approx 150$ MHz, corresponds to constraining $\sigma_l \gg 2D$, where $D$ is the diameter of an array tile.
Breaking this condition would require quite a peaked baseline density indeed.



With this in mind, for an arbitrary radially smooth layout, the solution is of the form
\begin{equation}
W_\nu = g(q)/f^2.
\end{equation}
We can use the same procedure to determine the final integral of $I$ (except that it has an extra factor of $2\pi i \vect{l}\cdot\vect{u}'_i$ in the exponent). 
This implies that the factors of $f^2$ cancel, so that we have
\begin{equation}
I =  \frac{g'(\vect{u}, \vect{l})}{g(\vect{u})} \int df \phi_\nu e^{-2\pi if\omega},
\end{equation}
and it is clear that the solution must be separable in $u$ and $\omega$. 
This clearly defines a ``brick'' structure valid for all $u$ (and which again has a cut-off at $\tau/\sqrt{2}\pi$).
Thus a wedge is precluded for any radially smooth layout.

We note that this was determined for arbitrary angular density $\rho_\theta$. 

\section{Description of Numerical Techniques}
	\label{app:numerical}
	The gridding of visibility data and its transformation into an averaged power spectrum are non-trivial tasks that can require a significant computational effort.
	In this appendix we describe the simple method we have taken in this work to accelerate this process and ensure its accuracy.
	
	Naively, the application of Gaussian weights from each baseline $\vect{u}'_j$ to a particular point of interest $\vect{u}_i$ is an order $N\times M$ calculation (where $N$ is the number of baselines, and $M$ the number of grid-points at which to evaluate the power spectrum).
	Several standard algorithms can reduce this calculation to of order $N\log M$.
	The most popular is to use an FFT-backed convolution.
	However, we do not choose this route, as it requires the  $\vect{u}_i$ to be arranged on a regular Cartesian grid, which has its own difficulties in terms of angular averages and dynamic range.
	
	Instead, we use a KD-tree algorithm (from the \textsc{scikit-learn} Python package) to efficiently determine the baselines within a given radius of every $\vect{u}_i$, and apply the weights
	from only these baselines. 
	The radius can be arbitrarily set, based on the beam width.
	This allows the $\vect{u}_i$ to be placed arbitrarily. 
	Since we require an angular average, it is most convenient to choose the $\vect{u}_i$ in a polar grid, so that the angular average is merely the average of a particular row in the array. 
	This has the dual benefits of simplicity and accuracy -- the average is specified at a particular magnitude of $q$, rather than an average over a complicated distribution of $q$ within an annulus. 
	
	This algorithm enables the numerical calculation of the 2D PS as an arbitrarily precise quantity.
	That is, if the number of nodes in an angular ring is arbitrarily large, the operation exactly converges to the integral Eq. \ref{eq:power_general}.
	In practice then, if one simultaneously tests for convergence, this algorithm provides an exact non-gridding solution to the numerical calculation of the 2D PS.
	In this paper we do not formally test for convergence, but rather simply use a number of angular nodes we deem to be sufficient to capture the integral adequately.
	In real-world applications, the extension to formal convergence-monitoring is rather simple, and may provide for quite efficient accurate calculations of the 2D PS.
	
\end{appendix}

\bibliographystyle{apj}
\bibliography{library}

\end{document}